\documentclass[
 reprint,
 amsmath,
 pre,
 amssymb,
 nofootinbib,
 aps,
 longbibliography,
]{revtex4-1}
\usepackage[pdftex]{graphicx}
\usepackage{wasysym}
\usepackage{amsfonts}
\usepackage{ifsym}
\usepackage{pifont}
 \usepackage{footmisc}
\usepackage[export]{adjustbox}
\usepackage{dsfont}
\usepackage{lipsum}

\let\oldaddcontentsline\addcontentsline

\newcommand{\starttocentries}{\let\addcontentsline\oldaddcontentsline}
\makeatletter
\newlength{\apb@width}
\newcommand{\autoparbox}[2][c]{\settowidth{\apb@width}{#2}\parbox[#1]{\apb@width}{#2}}

\makeatother

\newcommand{\namedref}[2]{\hyperref[#2]{#1~\ref*{#2}}}

\newcommand{\Tr}{\operatorname{Tr}}

\renewcommand{\Re}{\mathop{\mathrm{Re}}}

\newcommand{\Csphere}{{}^\bullet\kern-1.2pt C}
\newcommand{\Ctorus}{{}^\circ\kern-1.2pt C}

\newcommand{\nn}{\nonumber}

\newcommand{\COMMENT}[1]{}

\newcommand{\neqa}{\nonumber\end{eqnarray}}
\newcommand{\la}[1]{\label{#1}}

\newcommand{\<}{{\langle}}
\renewcommand{\>}{{\rangle}}

\newcommand{\re}{\relax{\rm I\kern-.18em R}}

\def\su2{{SU(2)}}

\def\a{{\alpha}}

\def\[{\left[}
\def\]{\right]}

\def\a{\alpha}

\def\({\left(}
\def\){\right)}
\def\[{\left[}
\def\]{\right]}

\def\<{\langle}
\def\>{\rangle}

\def\i2{\frac{i}{2}}

\def\cO{{\cal O}}

\def\2F1{\,_2{\rm F}_1}

\usepackage[usenames,dvipsnames]{xcolor}
\usepackage{color}
\usepackage{graphicx}
\usepackage{dcolumn}
\usepackage{bm}

\usepackage{cancel}
\usepackage{ctable}
\usepackage{booktabs}
\newcolumntype{L}[1]{>{\raggedright\let\newline\\\arraybackslash\hspace{0pt}}m{#1}}
\newcolumntype{C}[1]{>{\centering\let\newline\\\arraybackslash\hspace{0pt}}m{#1}}
\newcolumntype{R}[1]{>{\raggedleft\let\newline\\\arraybackslash\hspace{0pt}}m{#1}}

\newcommand{\beq}{\begin{equation}}
\newcommand{\eeq}{\end{equation}}
\newcommand{\beqq}{\begin{equation*}}
\newcommand{\eeqq}{\end{equation*}}
\newcommand\beqa{\begin{eqnarray}}
\newcommand\eeqa{\end{eqnarray}}
\newcommand\beqaa{\begin{eqnarray*}}
\newcommand\eeqaa{\end{eqnarray*}}
\newcommand\bea{\begin{array}}
\newcommand\eea{\end{array}}
\newcommand\De{\Delta}

\usepackage[hidelinks]{hyperref}

\begin{document}


\title{
Complex Spin: The Missing Zeroes and Newton's Dark Magic
}

\author{Alexandre Homrich$^{a,b}$, David Simmons-Duffin$^{c}$, Pedro Vieira$^{b,d}$} 
\affiliation{\vspace{0.1cm} $^{a}$Department of Physics and Astronomy, University of Waterloo, Waterloo, Ontario, N2L 3G1, Canada} 
\affiliation{$^{b}$ Perimeter Institute for Theoretical Physics, 31 Caroline St N Waterloo, Ontario N2L 2Y5, Canada}
\affiliation{$^{c}$ Walter Burke Institute for Theoretical Physics, Caltech, Pasadena, California 91125, USA}
\affiliation{$^{d}$Instituto de F\'isica Te\'orica, UNESP, ICTP South American Institute for Fundamental Research, Rua Dr Bento Teobaldo Ferraz 271, 01140-070, S\~ao Paulo, Brazil}

\begin{abstract}
Conformal Regge theory predicts the existence of analytically continued CFT data for complex spin. How could this work when there are so many more operators with large spin compared to small spin? Using planar N=4 SYM as a testground we find a simple physical picture. Operators do organize themselves into analytic families but the continuation of the higher families 
have zeroes in their structure OPE constants for lower integer spins. They thus decouple. Newton's interpolation series technique is perfectly suited to this physical problem and will allow us to explore the right complex spin half-plane. 
\end{abstract}

\pacs{Valid PACS appear here}

\hfill CALT-TH 2022-038

\maketitle

\section{Introduction}

Certain CFT data can be analytically continued to complex values of spin $S$ \cite{Caron-Huot:2017vep}. Specifically, for each four-point function~$\langle \mathcal{O}_1 \mathcal{O}_2\mathcal{O}_3\mathcal{O}_4\rangle$, Caron Huot's Lorentzian inversion formula produces functions $c^\pm_{1234}(\Delta,S)$ with the following properties. For integer $S$, they have simple poles at the locations of local spin-$S$ operators $\cO_i$ appearing in the OPE~$\cO_1\cO_2\to \cO_3\cO_4$:
\beq
c^{(-1)^S}_{1234}(\Delta,S) \sim \sum_i -\frac{f_{12i}f_{34i}}{\De-\De_i}, \quad(S\in \mathbb{Z}),
\eeq
where $f_{12i},f_{34i}$ are OPE coefficients.\footnote{In other words, $c^{\pm}_{1234}$ give analytic continuations of CFT data away from even-spin/odd-spin.}
Furthermore, the~$c^\pm_{1234}(\De,S)$ are analytic in $S$ and bounded for~$\Re(S)>1$, when $\De=\frac{d}{2}+i\nu$ is on the principal series. The latter condition is crucially tied to the boundedness of correlators in the Regge limit and conformal Regge theory \cite{Costa:2012cb,Caron-Huot:2017vep}.

Because the same local operators appear in every OPE (modulo global symmetries), different $c(\De,J)$-functions associated to different four-point functions have identical pole locations at integer $S$. We call this ``local operator universality."

\begin{figure}[t]
\centering
\includegraphics[trim={3cm 0 0 0},width=0.48\textwidth]{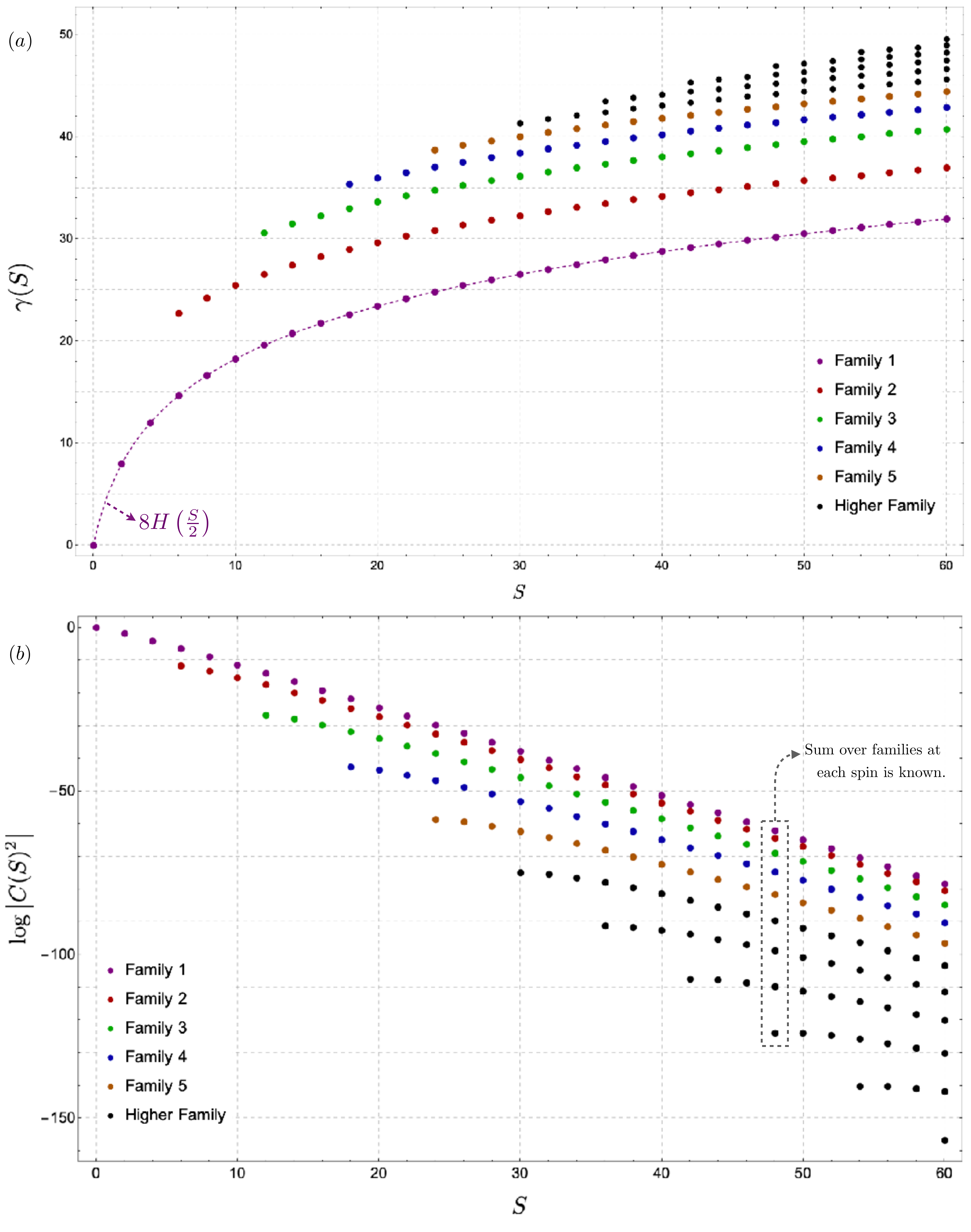}
\vspace{-0.3cm}
\caption{(a) Leading order anomalous dimensions $\gamma(S)$ where~$\Delta(S)=3+S+g^2 \gamma(S)$ and (b) square of structure constants~$C(S)^2$ for the twist $3$ operators of even spin. The energy of the lowest trajectory (\ref{energyLowest}) as well as sum rules of all structure constants for fixed $S$ (\ref{sum}) are known for any complex spin $S$; the structure constants for the lowest family~(\ref{C123lowest}) are derived here for even integer $S$. }
\label{twist3ope}
\end{figure}

What happens at non-integer $S$ is less clear. In \cite{Kravchuk:2018htv}, it was argued that singularities in $\De$ at non-integer $S$ should be interpreted as matrix elements of light-ray operators. However, little is known about the structure of such singularities. Are there always single poles in $\De$, or can there be higher poles, or cuts? Is there ``light-ray operator universality," where the same light-ray operators appear in different $c(\De,S)$ functions, or do different four-point functions see completely different light-ray operators?

Perhaps the simplest hypothesis is that there is ``light-ray operator universality" --- that is, local operators live on discretely-spaced Regge trajectories, whose dimensions $\De(S)$ and structure constants $C(S)$ are analytic in $S$.\footnote{Though there will be accumulation points in twist-space~\cite{Fitzpatrick:2012yx,Komargodski:2012ek}.}  Furthermore, one should be able to construct the corresponding light-ray operators directly, without reference to a particular four-point function or $c_{1234}(\Delta,S)$ that they appear in. Some evidence for this idea can be found in perturbative constructions of certain classes of light-ray operators \cite{Christ:1972ms,Gross:1973ju,Georgi:1974wnj,Collins:1981uw,Balitsky:1987bk,Kuraev:1977fs,Balitsky:1978ic,Mueller:1994jq,Balitsky:1995ub,Caron-Huot:2013fea,Caron-Huot:2022eqs}.

\underline{However, this idea raises an obvious puzzle:}

There are usually many more local operators with large spin $S$ than with small spin $S$. We illustrate this in figure~\ref{twist3ope}, where we plot the leading-order dimensions and  structure constants of twist $3$ operators in planar $\mathcal{N}=4$ SYM with even spin (up to spin sixty). As these figures clearly illustrate, the operators do seem to be nicely organized into smooth trajectories --- or families ---  which we made more visible with colouring. This is consistent with analyticity in spin mentioned above. On the other hand, the contribution of the various families to physical observables is ostensibly non-analytic, given the disappearance of physical operators at small spin. How can these two facts be reconciled?

\underline{There is a simple possible solution to this puzzle:}

Perhaps the analytically continued structure constants~$C(S)$ of a higher family which starts at some large physical $S_*$ will have zeroes at all integer spins below $S_*$, where a local operator is ``missing". In other words, an infinite number of Regge trajectories of light-ray operators exist for complex $S$, but most of them have zeros at integer spins.\footnote{
In \cite{BaltMadalena} Regge trajectories of the $(2,0)$ theory were studied. Several related puzzles were raised there and -- for some of those (see \textit{issue 2} there for instance) -- one of the proposed decoupling mechanisms is identical to the one observed here.  Could the Newton methods employed here  help sharpen some of the conjectures in \cite{BaltMadalena}?
}

In this letter we confirm this picture. Using an extrapolation technique found in Newton's 1687 \textit{Principia} we will continue physical\footnote{We use \textit{physical} to refer to the data directly related to local operators. The continuation of the data is as physical as the local data once we consider null Wilson line operators \cite{Kravchuk:2018htv,Balitsky:1987bk,Balitsky:2013npa,Braun:2003rp}.} data at integer spins into the complex plane. What we will see can only be described as dark magic: these Newton series extrapolations will beautifully converge to functions with precisely these zeroes at the locations of ``missing" operators. 

\section{The Twist 3 Data}
Figure \ref{twist3ope}a contains the one loop anomalous dimensions~$\gamma(S)$ for all primary operators of the form 
\beq
\Tr(D_+^S Z^3)+ \texttt{permutations} \la{family}
\eeq
with even\footnote{We will consider even spin for the most part. It is well known that analytic continuation of even and odd spins are independent. We checked indeed that similar conclusions would be reached for odd spins.} spin $S$. These operators have three units of R-charge and classical twist equal to three. Figure \ref{twist3ope}b depicts the three point functions $C(S)$ between these operators and two scalar BPS operators.\footnote{Any scalar BPS operators would lead to the same $S$-dependence of the three point function. These BPS operators are scalars with protected dimensions. We will not discuss them  further.} 

The family of operators (\ref{family}) will be central player in this note. These operators are the \textit{second} simplest spinning operators in this conformal gauge theory. The simplest would be the twist two operators -- simply replace~$Z^3$ by~$Z^2$ in (\ref{family}) -- which are however too simple for our purpose since there is a single lonely primary twist-two operator for each (even) spin $S$. The twist-two family is still very interesting, as reviewed in appendix \ref{twist2}. 

\begin{figure}[t]
\centering
        \includegraphics[scale=0.33]{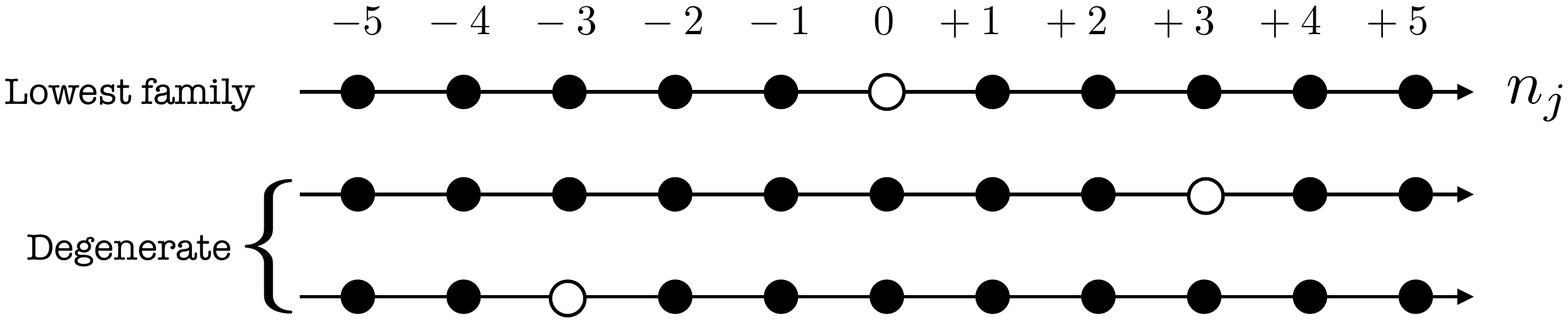}
    \vspace{-5cm}
\caption{The $u_j$'s need to be all distinct since these parametrize the momentum of the $S$ excitations in the spin chain which behave as fermions.  
Since $F$ and $f$ vanish at the origin, solutions with distinct $u_j$ require distinct external fields $n_j$. Since $F$ and $f$ are bounded functions these mode numbers must lie in the range quoted in the main text and since there are $S+1$ integers in this range we see that one of them -- called the \textit{hole} $n_*$ -- is not used. There is one additional selection rule on this $n_*$: It must generate a state of zero momentum (this is a gauge theory condition: the trace in (\ref{family}) means we are only interested in cyclic symmetric quantum spin chain states). This requires $n_*/3$ to be an integer. There are $2 \lfloor S/6\rfloor+1$ such choices which perfectly matches the counting of primaries. In the figure we illustrate the two possible choices for spin $10$ corresponding to the lowest family and the second family; the black dots are the $n_j$ and the white circle is the hole $n_*$.}
\label{holeFigure}
\end{figure}

At weak coupling, the primary operators (\ref{family}) can be thought of as eigenstates of an SL$(2,R)$ quantum spin chain of length $3$ with $S$ excitations, the conformal dimension of the operators corresponding to the energy levels of the quantum spin chain \cite{Beisert:2004ry,Staudacher:2004tk}. This energy, in turn, can be determined by solving a simple electrostatic problem of $S$ charges in a line with real positions $u_j$ given by 
\beq
\sum_{j\neq k} f(u_j-u_k)+F(u_j) - 2\pi n_j=0, \label{electro} 
\eeq
where the \textit{external force} $F(u)=6 \arctan(2u)$, the \textit{interaction forces} $f=2\arctan(u)$ and the mode numbers\footnote{In the electrostatic picture we can think of them as an extra constant force felt by each particle. We can think that the particles have charge $n_j$ and are acted by a constant electric field of magnitude $2\pi$ for instance to explain the last term in (\ref{electro}).} $n_j$ are distinct integers taking all values from $-S/2$ to $S/2$, skipping one value which we call $n_*$. The twist three families depend on this missing mode number $n_*$ also denoted as \textit{the hole}, see figure \ref{holeFigure}.
The lowest (dimension) family has~$n_*=0$.\footnote{Dimensions are monotonic as a function of $|n_*|$.} All other families are degenerate pairs with $n_* = \pm1, \pm2 ,\dots$. 

Solving the electrostatic equations (\ref{electro}) is trivial: \texttt{Mathematica}'s built-in \texttt{FindRoot} does the job.\footnote{We solved these equations all the way to spin $200$ for all possible choices of $n_*$. We had to solve them with huge precision to make sure the corresponding energies and structure constants are accurately predicted. We found it very efficient to find the position of the particles $u_j$ first with some reasonable precision (using \texttt{FindRoot} with \texttt{WorkingPrecision->100} digits say) and then using these locations as a starting point to solve the equations again with more precision (using \texttt{FindRoot} again with \texttt{WorkingPrecision->1000} digits say)}
Once we find a $\{u_1,\dots,u_S\}$ we extract the energy -- that is the dimension of the corresponding operator -- by adding up the energies of each particle 
\beq
\Delta=3+S+g^2 \Big( \gamma \equiv \sum_{j=1}^S \frac{2}{u_j^2+1/4}\Big) \, . \la{energy}
\eeq
This is how figure \ref{twist3ope}a was generated. Similarly, the structure constant $C(S)$ is also given by an expression in terms of the $u_j$ from which we generated figure \ref{twist3ope}b. The explicit plug-in expressions are simple but not as simple the spectrum expression: $C(S)$ is given by a ratio of  determinants of matrices of size $S$ built out of these $u_j$ or equivalently in terms of integrals of so called Baxter polynomials $Q(x)=\prod_{j=1}^S (x-u_j)$, as reviewed in appendix~\ref{Cappendix}. 

Of course, this integrability description of the conformal data is tailored for physical operators: It is hard to put a non-integer number of particles on a line or make sense of determinants of matrices with non-integers sizes! We will soon discuss how to (partially) overcome this limitation.

This concludes the description of how we obtained all the physical twist three data. There are a few analytic things we know about it. 

For the lowest family, we know $\Delta$ and $S$ analytically for any even spin $S$. They are rational numbers. For even~$S$, we have 
\beqa
\Delta(S)_\texttt{lowest family} &=& 3+S+8 g^2 H(S/2) \label{energyLowest}\\ 
C(S)_\texttt{lowest family}^2 &=&  \frac{4 S!^2}{(2S+1)!}  \times \label{C123lowest} \\
&&\!\!\!\!\!\!\! \!\!\!\!\!\!\! \!\!\!\!\!\!\! \!\!\!\!\!\!\! \!\!\!\!\!\!\!\!\!\!\!\!\!\! \!\!\!\!\!\!\! \!\!\!\!\!\!\! \times\!\left(\sum_{j=0}^{\infty}\!  \left(\! \frac{ \sqrt{1+4j}\,\Gamma\big(\tfrac{1}{2} +  j\big)^{\!2} \Gamma\!\left(1+\frac{S}{2}+j\right)\!/\Gamma(j+1)^2}{ \Gamma\!\left(\tfrac{1}{2}-\tfrac{S}{2}+j\right) \Gamma\!\left(1+\tfrac{S}{2}-j\right) \Gamma\!\left(\tfrac{3}{2} +\frac{S}{2}+j\right)} \right)^{\!\!2}\right)^{\!\!-1}  \!\!\!\! . \nn
\eeqa
The expression (\ref{energyLowest}) for the energy of this lowest family was first found in \cite{Beisert:2004ry,Beccaria:2007cn}, see also appendix \ref{lowestFamilyAppendix}. 

The expression (\ref{C123lowest}) for the structure constant is new. It is derived in appendix~\ref{lowestFamilyAppendix} using the recently developed SoV representation of structure constants in this gauge theory \cite{Bercini:2022jxo}. We can use it to very efficiently produce physical data for spins as large as we want.\footnote{Note that the sum in the second line of (\ref{C123lowest}) can be truncated to~$j\le S/2$ since all terms with bigger $j$ vanish. This sum in (\ref{C123lowest}) can actually be trivially done: the result is a sum of two $_{8}F_{7}$'s.} Evaluating~(\ref{C123lowest}) for $S=10$, for instance, leads to~$C(10)_\texttt{lowest family}=1/84756$. 

There is something unusual about (\ref{C123lowest}): this expression holds for all even integers $S$ -- there is nothing wrong with it if we stick to physical operators -- and it looks perfectly analytic in spin and yet it is \textit{not} the correct analytic continuation of the physical data to complex spin $S$. Indeed, if we were to study this function in the complex plane we would find a myriad of unphysical poles in the right half-plane (RHP) (from zeroes of the big parentheses in the last line), violating the expected Regge bahavior of this theory. 
The \textit{correct} analytic continuation away from the even integers $S$ is given to us by Newton as explained in the next section. 

We also know
\footnote{We can derive this result in two trivial ways. We can construct this sum for various spins $S$ from the solutions of the electrostatic problem and simply observe that they follow a nice sequence given by this expression; \texttt{Mathematica}'s \texttt{FindSequenceFunction} would give it right away. Alternatively we could simply decompose the leading twist contribution to the tree level BPS correlator (take (66) from~\cite{Vieira:2013wya} say) \beqa G^\texttt{tree}_{2233}(z,\bar z)&=&\frac{\left(2 z \left(\bar{z}-1\right)-2 \bar{z}+3\right) \left(z \bar{z}\right)^{3/2}}{(z-1)^2 \left(\bar{z}-1\right)^2}\simeq \frac{(3-2 z) z^{3/2} \bar{z}^{3/2}}{(z-1)^2} \nn \\ \nn &&\!\!\!\!\!\! \!\!\!\!\!\! \!\!\!\!\!\! \!\!\!\!\!\! \!\!\!\!\!\! = \bar{z}^{3/2} \sum\limits_S \frac{\texttt{sum}(S)}{C_{233}^2} z^{S+\frac{3}{2}} \, _2F_1(S+2,S+2;2 S+3;z)\eeqa to read off (\ref{sum}) for even spin and $\texttt{sum}(S)=S!^2/(2S+1)! \times (S+1)(S-1)/3$ for odd spin. From a four point functions point of view we can only go beyond extracting such sum rules if we explore higher loops and/or multiple correlators otherwise all we access are these sums over classically degenerate operators.} 
analytically the sum of all $C(S)^2$ is for each spin $S$. These sum rules appear in correlation functions, where all operators appear in the OPE. They evaluate to nice rational numbers \cite{Vieira:2013wya}. Here, we find 
\beq
\!\!\texttt{sum}(S)\equiv\!\! \sum_\texttt{families} \!\! C(S)^2 = \frac{S!^2}{(2S+1)!}\! \times\! \frac{(S+1)(S+3)}{3}\,.   \la{sum}
\eeq

For spin $S=10$, for instance, we get~$\texttt{sum}(10)=1/81396$. At this spin, there are only two families and since we know the lowest one analytically we can read off the second family as\footnote{Recall that all higher families are double degenerate so that $\texttt{sum}(10)$ is equal to $C(10)^2_\texttt{lowest family}+2\,C(10)^2_\texttt{second family}$.} $C(10)^2_\texttt{second family}=5/20532141$. 
Above spin $S\ge 12$, we have more than two families and only the lowest family is given by a rational number. All other solutions are given by more complicated algebraic numbers which we computed numerically with several hundreds of digits for all spins up to $S=200$.

This concludes the study of the twist $3$ data for physical integer spin. We now need to go to the complex spin plane. Enter Newton. 

\section{Newton's Magic}

Given the CFT data at spins $S=2,4,6,\dots$, we want to go to the complex plane. Normally this is not a well posed problem; we can always add a bunch of simple (trigonometric) functions that vanish at the integers to any possible analytic continuation. 

\begin{figure}[t]
        \includegraphics[width=0.48\textwidth]{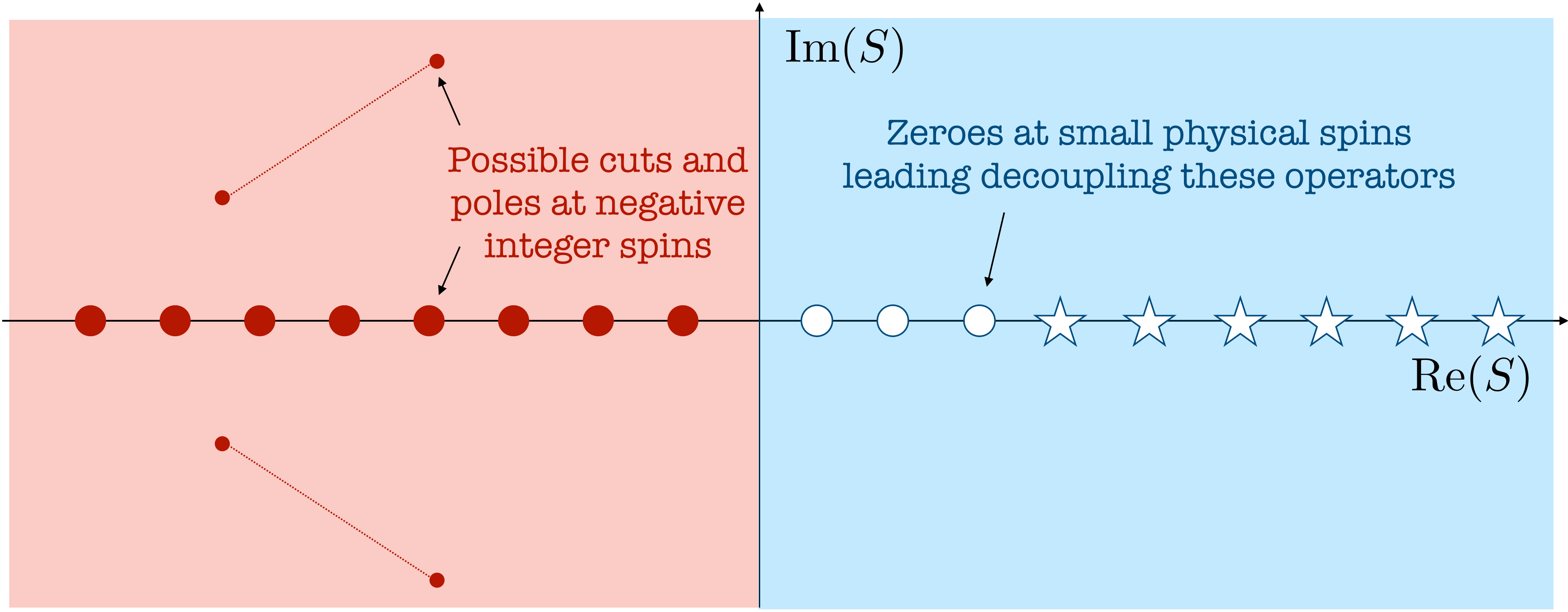}
        \vspace{-0.2cm}
\caption{Cartoon for complex $S$ plane behavior for $C(S)$. The main result of this note is the confirmation of the expected zeroes in the right half-plane.}
\label{C123complexPlane}
\end{figure}

Here, however, physics come to the rescue to render the continuation unique. Conformal Regge theory ensures that the $c(\De,S)$ functions discussed in the introduction should be bounded and free of singularities in the complex $S$ plane to the right of $S=1$ \cite{Costa:2012cb,Caron-Huot:2017vep}. In a planar theory, this requirement is slightly weakened: singularities are excluded to the right of $S=2$, as a consequence of the bound on chaos \cite{Maldacena:2015waa}. We hypothesize that this requirement extends to the individual structure constants $C(S)$ of each Regge trajectory (which are residues of $c(\De,S)$ at the locations of poles). A rough argument is that different $C(S)$ can be accessed as different residues of $c(\De,S)$ in $\De$. As long as the $\De$ are not degenerate, these residues should be individually finite\footnote{We verify no such degeneracies occur in the RHP through an integrability based method in section \ref{BaxterBlessing}. That the method presented in this section works suggests in itself the absence of degeneracies, as discussed in the main text.}. Finally, in our case, supersymmetry relates the twist-3 operators of interest to operators with higher spin by 2 units\footnote{ See \cite{Dolan:2001tt, Costa:2013zra} for the precise Ward-identities in the twist 2 case.}. Thus, we expect that $C(S)$ should be bounded and free of singularities to the right of $S_*=0$.

 (At the same time, in the left half plane we do expect a host of singularities of various types as depicted in figure~\ref{C123complexPlane}.)

Together, these conditions suffice for Carlson's theorem, which ensures the uniqueness of the analytic continuation of this data.\footnote{Carlson would require less. The function could grow exponentially along the imaginary direction provided the exponent is smaller than $\pi$ (this is roughly speaking what rules out trigonometric functions like $\sin$'s and $\cos$'s while $\tan$'s and $\cot$'s are excluded by the absence of singularities.) } 

What is perhaps less well-known is that the unique extension alluded to in Carlson's theorem can be explicitly constructed by a beautiful interpolation series written down by Newton in 1687's \textit{Principia Mathematica}.\footnote{It is Case 1 of Lemma V of Book III on the problem of \textit{To find a curve line of the parabolic kind which shall pass though any given number of points}.} Newton's series  
\beq
f_N(z) \equiv \sum_{j=0}^N \binom{z}{j} \sum_{i=0}^j \binom{j}{i}(-1)^{j-i}f(i)\,. \label{Newton}
\eeq
converges to the proper extension $f(z)$ as $N\to \infty$. 
For us,~$f$ could be either the energy of the structure constants and the argument would be $z=S/2-S_n/2$ where $S_n$ is the first value for which a given family exists. That is is~$S_1=0$ for the lowest family,~$S_2=6$ for the second family,~$S_3=12$ for the third family etc. Then physical values correspond to $z=0,1,2,\dots$ and we include $N$ of these in the interpolation. Newton's series is very powerful: it supposed to converge to the right of the first allowed singularity, so for us it should converge in the full right half complex spin plane \cite{zbMATH02586461}.

\begin{figure}[t]
  \includegraphics[width=0.48\textwidth]{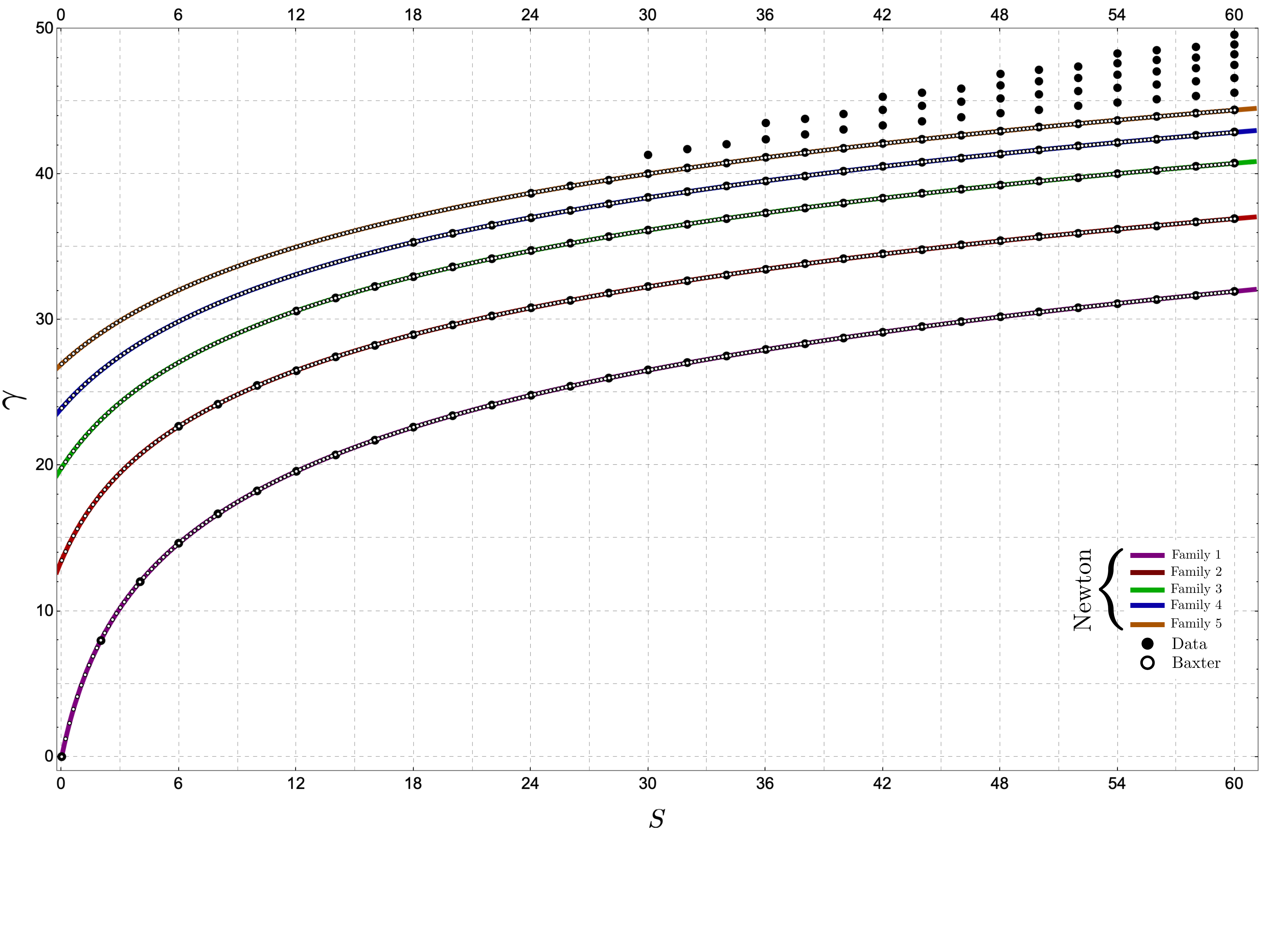}
  \vspace{-1.2cm}\caption{The Newton continuation of the operators dimensions is represented here for the lowest five families at $N=88$. The white dots corresponding to an alternative integrability method described in section \ref{BaxterBlessing}. They lie perfectly on top of 
  the Newton continuations.}  \la{energiesfig}
\end{figure} 

\begin{figure*}[t]
  \includegraphics[width=\textwidth]{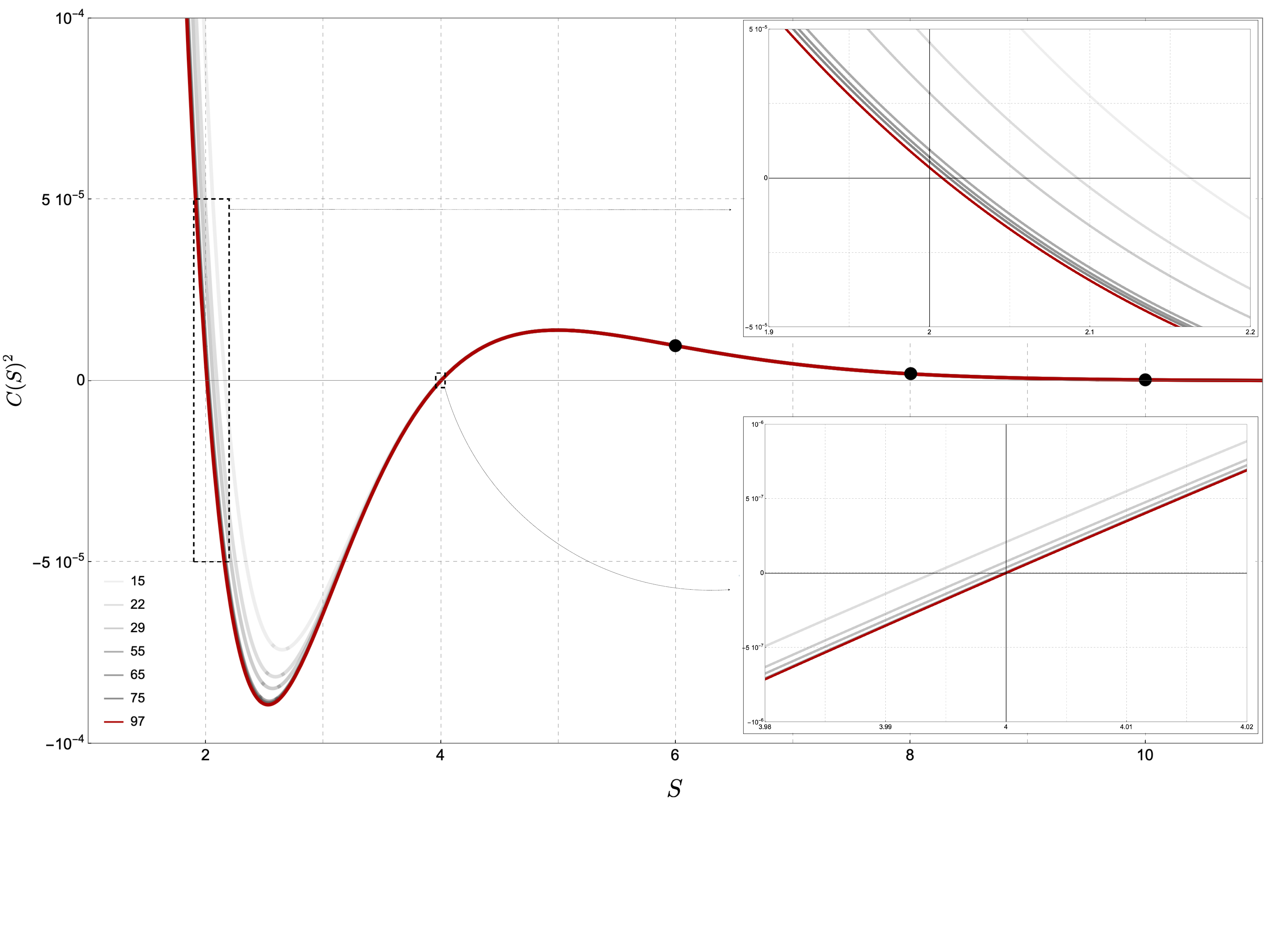}
  \vspace{-2.5cm}
  \caption{Newton's continuation of the structure constants for the second family converging as we include more and physical operators for $N=15,\dots,97$. The outcome of Newton's interpolation is almost spooky: The physical structure constants are \textit{growing} as we come from the right approaching $S=6$ and the continuation reverses this growth and dives perfectly through zero surgically at (the even integer) $4$ coming back up through zero again precisely at (the even integer) $2$!  If this is not dark magic, what is dark magic? } \label{DarkMagicFigure}
\end{figure*}

Applying this method to the anomalous dimensions of the various trajectories with $N=88$ results in figure \ref{energiesfig}. Experimentally, the method quickly converges as $N$ is increased. This confirms the picture that, indeed, each trajectory $\gamma(S)$ is free of singularities in the RHP, as otherwise the method would not have converged. Of course, given convergence, one could simply scan over the complex plane and observe that no degeneracies occur away from the real line. In section \ref{BaxterBlessing} we discuss an alternative integrability-based method that allows for the computation of the energies directly at complex spin. As shown in figure \ref{energiesfig}, it confirms the absence of operator mixing in the RHP and the numerical accuracy of the Newton series.

Having confirmed the lack of degeneracies, let us now jump to the most exciting application of this method: The study of the structure constants for the second family, which starts at spin $S=6$. We first apply~(\ref{Newton}) using the first~$N=15$ physical spins; then we include more and more physical operators in the interpolation up to~$N=97$, see figure \ref{DarkMagicFigure}. As we increase $N$ we see that the analytic continuation beautifully converges\footnote{To be precise, we continue $2^{\log(2) S} C(S)$ and later subtract the exponential. The reason for this is that the structure constants decay \textit{too quickly}, which turns out to be an issue for the series convergence. Generically, once can always subtract the exponential behaviour of the interpolated sequence to obtain a convergent series representation provided there is no Stokes-like phenomena in the RHP. This is the case here due to Regge theory.} and precisely predicts zeros at the lower spins $S=4$ and~$S=2$ as mentioned in the introduction!

Repeating (a refinement of) this analysis for all families we end up with figure \ref{summaryallfamilies}.\footnote{We could in principle go all the way to $S=0$ but it is harder to get convergence there; this is of course not surprising as $S=0$ is the boundary of the RHP. We are agnostic about the expectation of (non) emergence zeroes at $S=0$.} The picture put forward in the introduction with the decoupling of the various higher families at lower physical spins is neatly realized.

\begin{figure}[t]
  \includegraphics[width=0.48\textwidth]{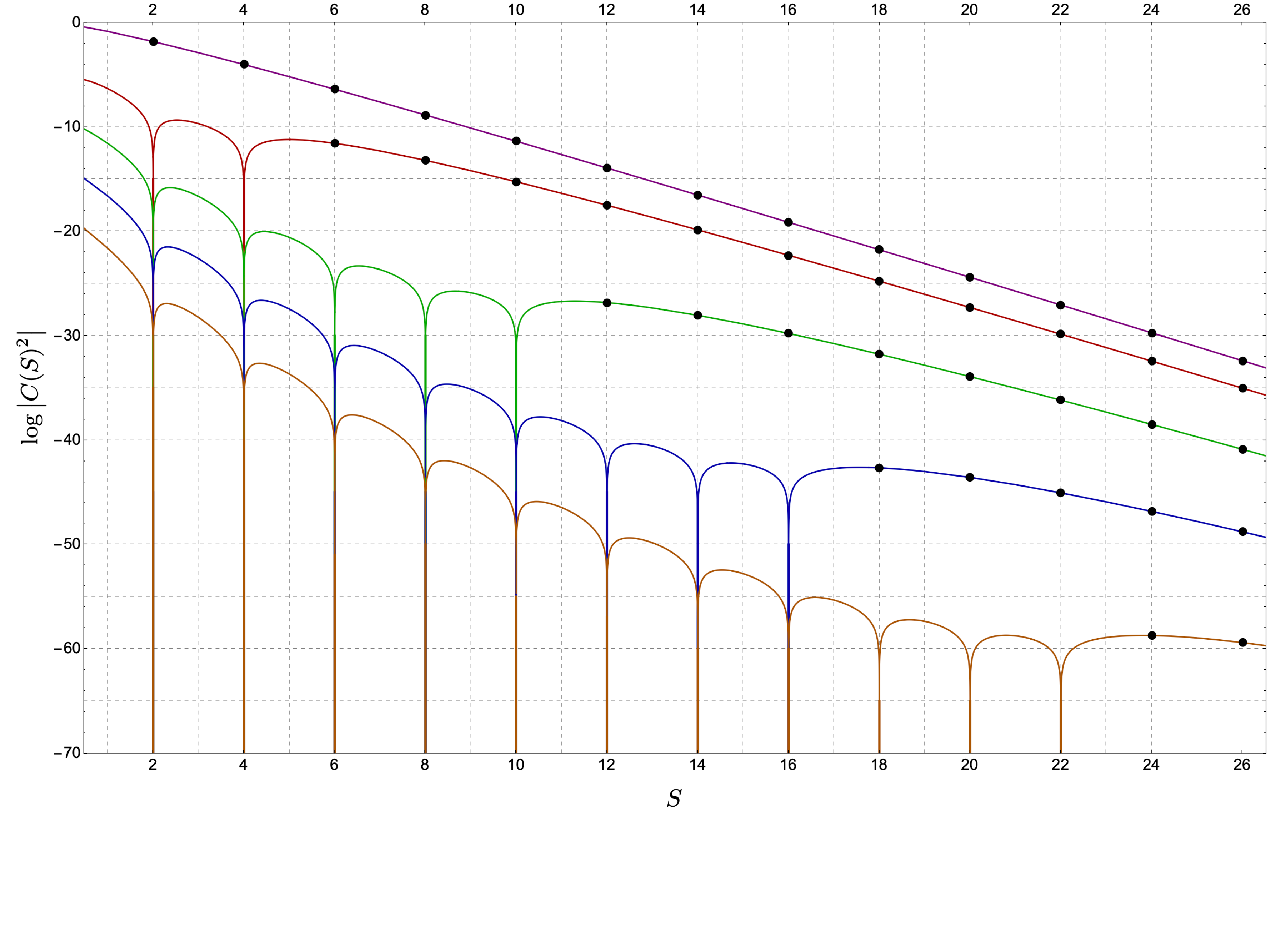}
\vspace{-1.3cm}
  \caption{To extend the analytic continuation for all $S>0$ we first run the interpolation for a family starting at $S_n$ and check that a zero emerges dynamically as $S_n-2$ as we increase~$N$. Once this happens we can put it in by hand and continue instead the Newton's series interpolation of~$C^2(S) \times (S-S_n+2)$ and see if a zero at $S_n-4$ emerges; If so we include it and interpolate $C^2(S) \times (S-S_n+2) \times (S-S_n+4)$ and so on, always checking for stability of the procedure.
  } \la{summaryallfamilies}
\end{figure}

\section{Baxter's Blessing} \label{BaxterBlessing}
Can we cross-check the analysis of the last section and compute directly the CFT data at complex spin? We do not know how to do it for structure constants -- as discussed below -- but for the spectrum $\gamma(S)$ Baxter equations give an alternative method we can use which reduces to the previous electrostatic description for integer spin, but generalizes neatly for complex spin. 

The starting point is the Baxter twist three equation
\beqa
Q(u+i)(u+i/2)^3+Q(u-i)(u-i/2)^3 \nn \\- Q(u)\big(2u^3-(S^2+2 S+\tfrac{3}{2}) u+q\big) = 0. \la{baxterEq}
\eeqa
For integer spin, this equation is equivalent to the electrostatic problem described above. Indeed, for integer $S$ we look for polynomial solutions which we parametrize as~$Q(u)=\prod_{j=1}^S (u-u_j)$. Evaluating this equation at~$u=u_j$ we find that the second line vanishes and thus the ratio of the two terms in the first line must be equal to~$-1$ when~$u=u_j$. The logarithm of this statement is the electrostatic equation (\ref{electro}); the integers $n_j$ correspond to the various possible log branches. 
From the Baxter point of view, the different trajectories are labelled through their $q \equiv q(S,n_*)$ eigenvalue, which is self-determined from (\ref{baxterEq}) plus polinomiality. 

What is beautiful about this equivalent Baxter formulation is that it now allows for an extension to complex spin $S$, once we relax the polynomiality property of the Baxter function $Q(u)$. 
The proposal is that we should look for the slowest growing entire solution of (\ref{baxterEq}). We review this prescription in appendix \ref{baxterAppendix}.

A numerical algorithm to obtain such solutions is as follows. Assume $Q$ is given by a truncated power series normalized at $u=i/2$: 
\beq
Q(u) = 1 + \sum_{k=1}^{N_\text{max}} a_k (u-i/2)^k. \la{ansatz}
\eeq
Plugging (\ref{ansatz}) in (\ref{baxterEq}) and requiring that the first $N_\text{max}$ orders of the expansion around $u=0$ are satisfied completely fixes the coefficients $a_k$ in terms of $S$ and $q(S, n_*)$. The claim is that this procedure, in which we approximate the power series tail by zero, converges to the slowest growing entire solution as $N_\text{max} \rightarrow \infty$. One is left with solving the quantization problem of determining $q(S,n_*)$ for complex S. This is achieved by requiring the cyclicity condition $Q(i/2) = Q(-i/2)$, see figure \ref{holeFigure}. 

\begin{figure}[t]
  \includegraphics[width=0.48\textwidth]{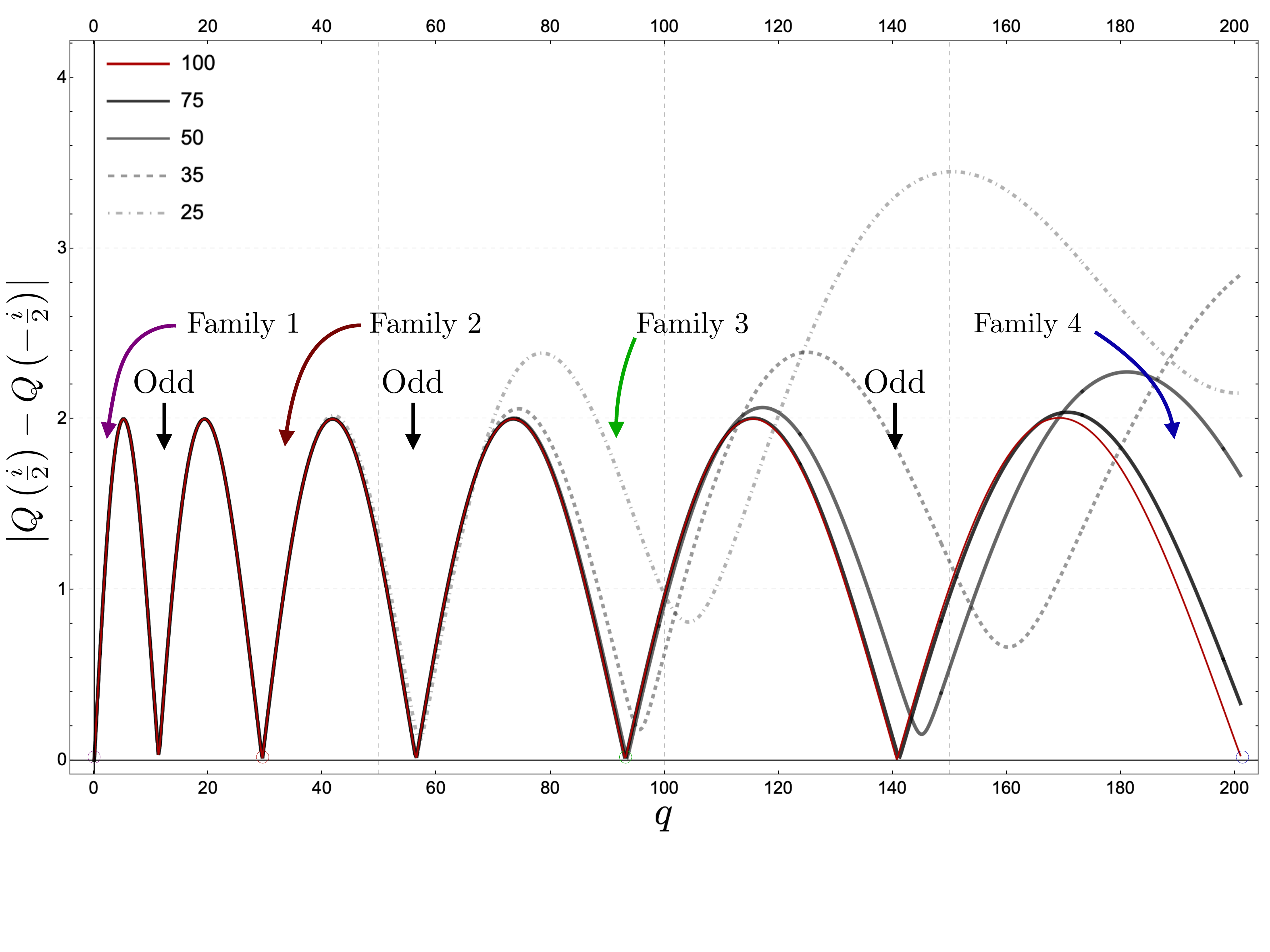}
\vspace{-1.2cm}
  \caption{
 We can find the physical values of $q$ by requiring that the continued Baxter solution preserves the zero momentum condition associated with single trace operators. More terms in the expansion (\ref{ansatz}) are needed to converge as we scan over higher values of $q$, corresponding to higher families, as indicated by the various  $N_\text{max}$ curves, which are evaluated at~$S = 11/3$ here; to find the third family $N_\text{max}=35$ is almost enough for instance while $N_\text{max}=50$ is definitely plenty. The method identifies both even and odd spin trajectories, which come as alternating zeroes, since the same Baxter equation describes both sectors. 
  } \la{quantization}
\end{figure}

This procedure is illustrated in figure \ref{quantization}. Starting from integer spin, one can identify which local minimum of~$|Q(i/2) - Q(-i/2)|$ corresponds to each family and adiabatically deform from there. Alternatively, one could determine $q(S, n_*)$ through the Newton's series of the physical data. Both methods show perfect agreement.

Having determined the power series (\ref{ansatz}) one can then simply extract the energies from 
\beq
\gamma = 2 i \left(Q'(\tfrac{i}{2}) - Q'(-\tfrac{i}{2})\right),
\eeq
which generalizes (\ref{energy}) to the non-polynomial case. The result for the first five families is presented in figure \ref{energiesfig}. It is in perfect agreement with the Newton's series continuation from the integer spin data.

\section{Discussion}
\la{discussion}

The complex spin plane is a formidable universe and planar $\mathcal{N}=4$ SYM, being a solvable higher dimensional conformal field theory is a perfect laboratory for its exploration.\footnote{The hydrogen atom is also a remarkable laboratory for complex spin studies as Regge highlighted; in appendix \ref{hydrogenAp} some hydrogen atom cute decouplings akin to those observed in this letter are presented.}

In this letter, we focused on the right half-plane~$\text{Re}(S)>0$. On this side, conformal Regge theory predicts controllable growth at large spin and no singularities at finite spin. As such, there is not much room for surprises in the RHP. The notable exception are the \textit{missing zeroes} we explored here. We found that the analytic continuation of structure constants for higher trajectories develops zeroes at integer positive spins, explaining their decoupling from correlation functions and resolving the puzzle raised in the introduction. 

We conjecture this decoupling picture to be a general mechanism for any higher dimensional conformal field theory. Of course, testing it is not easy. Here, the  data obtained in planar $\mathcal{N}=4$ SYM through integrability was precious. To convincingly observe these zeroes, we used about a hundred physical data points computed with hundreds of digits of precision and plugged those into a beautiful continuation formula by Newton. Of course, this is not a luxury we can afford in the 3D Ising model where we might have at most a few physical operators with a few digits of precision.

It would be very interesting to explore other theories where we might be able to compute large amounts of physical structure constants with great precision and look for the emergence of the missing zeroes there as well. We could also assume these zeroes to be there and use them to more efficiently continue physical data into the complex plane. 

In $\mathcal{N}=4$ SYM, we could also explore other sectors and learn about analyticity (or lack thereof) in several other parameters such as R-charges, mixed spins~$(S_1,S_2)$ for more complicated spinning operators, twist, etc. A very nice study of strong coupling double trace operator decoupling as one analytically continues their R-charge representations can be found in \cite{Aprile:2020mus}.

It is also an important open problem to understand how to directly compute the structure constants at any complex spin $S$ from integrability without resorting to Newton's magic -- as done for the spectrum in section~\ref{BaxterBlessing} using Baxter's Q-functions. In \cite{Bercini:2022jxo} new formulae were put forward for structure constants in this theory in terms of Separation of Variables (SoV) like integrals~\cite{Derkachov:2002tf, Cavaglia:2019pow, Jiang:2016ulr} of $Q$-functions so one might wonder if we can not simply plug the $Q$-functions found in the spectrum study there. Sadly, we do not know how to do it. Take for example the simplest possible leading order structure constant in the $SL(2)$ sector corresponding to twist-two operators. There is no operator degeneracy and a single family of operators in this case for which Dolan and Osborn \cite{Dolan:2000ut,Dolan:2003hv} extracted $C(S)^2_\texttt{twist-2 LO} = \frac{S!^2}{(2S)!}$ many years ago already. In integrability terms this can be obtained from~\cite{Bercini:2022jxo}
\beq
C(S)^2_\texttt{twist-2 LO} 
= \frac{S!^2}{(2S+1)!}\frac{1}{\displaystyle \int\limits_{-\infty}^{\infty}\!\! du\,  \frac{\pi}{2}\frac{Q(u)^2}{\cosh(\pi u)^2}} \,, 
\eeq
see appendix \ref{twist2}. For any even integer spin the $Q$ function is a polynomial so that integral is perfectly convergent; it evaluates to $1/(2S+1)$ \cite{Bercini:2022jxo}, see also appendix \ref{orthoapp} for a derivation. When $S$ is not integer, the Baxter function $Q$ grows as $e^{2\pi u}$ at infinity \cite{Janik:2013nqa} however and the integral no longer converges. It would be interesting to generalize~\cite{Bercini:2022jxo} so that it could apply not just for physical local operators of integer spin but also for analytically continued spins. One approach could be to try to make the SoV expressions manifestly gauge invariant under the symmetry $Q\to f_\texttt{i-periodic} \,Q$ of the Baxter equation.

A related question is whether we can explicitly write down the light-ray operators corresponding to the Regge trajectories we found in this work, in the spirit of \cite{Christ:1972ms,Gross:1973ju,Georgi:1974wnj,Collins:1981uw,Balitsky:1987bk,Kuraev:1977fs,Balitsky:1978ic,Mueller:1994jq,Balitsky:1995ub,Caron-Huot:2013fea,Caron-Huot:2022eqs}. Such light-ray operators should take the form
\beq
\label{eq:lightrayansatz}
\int d\a_1d\a_2 d\a_3 \psi(\a_1,\a_2,\a_3) \Tr(Z(\a_1)Z(\a_2)Z(\a_3)).
\eeq
Here, $Z(\a)$ is an insertion of $Z$ on a Wilson line lying along future null infinity, at retarded time $\a$. The three~$Z$ insertions are integrated against a wavefunction $\psi(\a_i)$, which must be translationally-invariant in retarded time to describe a primary light-ray operator. The homogeneity of $\psi$ is related to the spin $S$.\footnote{In the case of integer spin, $\psi$ should be a combination of $\delta$-functions and their derivatives, and (\ref{eq:lightrayansatz}) becomes the light-transform \cite{Kravchuk:2018htv} of a local operator.}

 In trying to directly construct such operators, we run into puzzles. Naively, at fixed $S$, the wavefunction $\psi$ can depend in an arbitrary way on the translationally-invariant homogeneity-zero combination $\chi=\a_{12}/\a_{13}$. In other words, there seem to be a {\it continuous infinity\/} of primary light-ray operators we can write down at each spin $S$, parametrized by an arbitrary function of $\chi$.
 \begin{figure}[t]
  \includegraphics[width=0.5\textwidth]{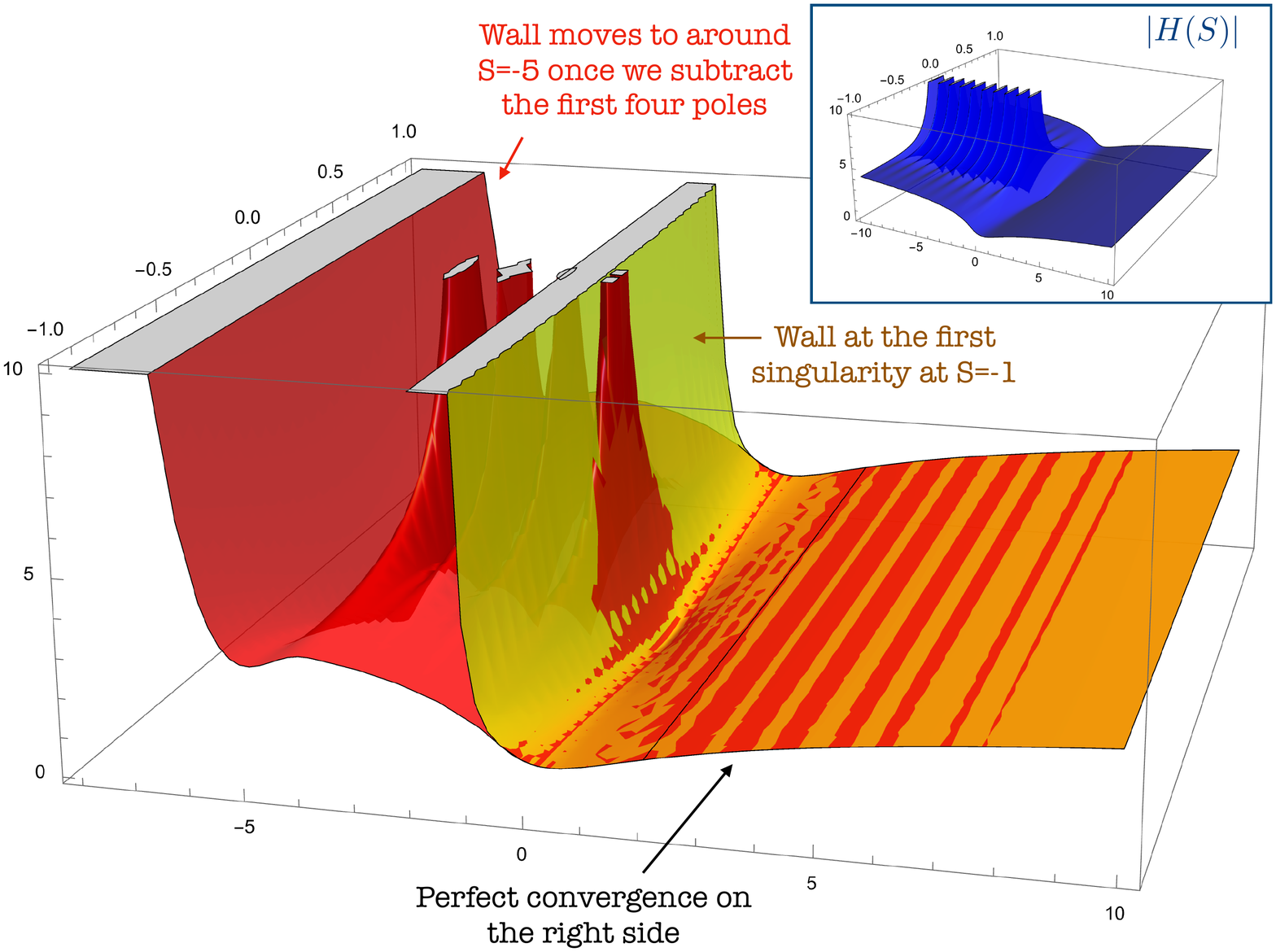}
  \caption{Newton continuation (\ref{Newton}) of $H(S)$ using $N=100$ integer values leads to the yellow curve (plotting the absolute value of the continuation). It perfectly approaches the analytic result (depicted in the inset) in the right-half plane (provided of course the spin is not too large) and will converge indeed perfectly in the full right-hand plane as we increase $N$. It stops converging, however, around~$S=-1$ where $H(S)$ has a pole. In this case, it is easy to tear down this wall of convergence. We simply continue $H(S)\times \big( p_k(S)\equiv (S+1)(S+2)\dots (S+k) \big)$ to write an improved continuation 
  \vspace{-0.2cm}
  \\ \\
  ${\color{white} em} \displaystyle f_N^\texttt{subtracted}(z)= \frac{1}{p_k(z)}  \sum_{j=0}^N \binom{z}{j} \sum_{i=0}^j \binom{j}{i}(-1)^{j-i} p_k(i) f(i)\,.$
    \\ \\
where $f(i)=H(i)$ in this example.
  With $k=4$ this leads to the red curve where we see that the perfect convergence is now extended to around $S=-5$.
  } \la{example}
\end{figure} 

 What reduces this continuous infinity of choices down to a discrete set of Regge trajectories? In our case, the mode number $n_*$ gave a natural way to define trajectories. This should translate into some natural conditions on the wavefunction $\psi$, and it would be nice to characterize them explicitly. Perhaps $\psi$ can be obtained from the analytically-continued $Q$-function.  
 Explicit expressions for the light-ray operators would hopefully manifest the missing zeros. A simplified toy model is the following wavefunction:
\beq
\frac{1}{\Gamma(-S)}|\alpha|^{S_*-S}.
\eeq
For integer $S>S_*$, the $\Gamma$-function in the denominator gives a zero that combines with the singularity in $|\alpha|^{S_*-S}$ to give a $\delta$-function derivative, which is appropriate for describing the light-transform of a local operator, see also appendix \ref{lightray}. However, for $S\leq S_*$, $|\alpha|^{S_*-S}$ has no singularity and the~$\Gamma$-function just makes the result vanish at the integers.

We can also ask: was our ability to cleanly define Regge trajectories an accident of single-trace operators of twist~3? To address this, it may be interesting to study analytic continuations of higher-twist families. Will such families enjoy unique analytic continuations, or might there be multiple ways to analytically continue them?\footnote{One might imagine different analytic continuations corresponding to different ways of folding a Wilson line like (\ref{eq:lightrayansatz}) with multiple insertions. We thank Simon Caron-Huot for this suggestion.}
 
 More generally, what mechanism could reduce the naive continuous infinity of light-ray operators (\ref{eq:lightrayansatz}) to a discrete set in non-integrable perturbative theories, like the Wilson-Fisher theory? Can we figure out the general mechanism by understanding operators in $\mathcal{N}=4$ SYM well-enough?

Everything we observed in this letter at leading order in perturbation theory should hold at finite coupling.\footnote{Confirming this picture is hard but it is not science fiction in planar $\mathcal{N}=4$ SYM. In \cite{Gromov:2015wca} and \cite{Alfimov:2014bwa} beautiful finite coupling analysis of the spectrum of twist two operators was carried out with spectacular results. Our twist three setup is more complicated and for structure constants we do not have the powerful technology of the quantum spectral curve at our disposal yet despite recent advances such as~\cite{Bercini:2022jxo} building on important previous works \cite{Derkachov:2002tf,Kazama:2013rya,Jiang:2016ulr,Giombi:2018qox,Gromov:2019wmz,Cavaglia:2019pow,Cavaglia:2021mft}.
}
Repeating the leading-order analysis at next-to-leading-order should be straightforward, for instance. More ambitiously, can the missing zeroes be useful in searching for a finite coupling version of \cite{Bercini:2022jxo}?

Finally, we have the left-half plane (LHP) of $S$, where there is a host of exciting physics to explore. In this left hand-side of the complex spin plane there will be dragons. In perturbation theory, we expect there a host of different singularities from poles at negative spins which would open up into cuts at finite coupling, creating an intricite infinitely-sheeted Riemann surface. This part of the plane contains so-called ``horizontal trajectories" of BFKL type \cite{Lipatov:1976zz,Kuraev:1977fs,Balitsky:1978ic} whose tree-level spin $S$ is fixed, but whose dimension $\De$ can vary. When the coupling is turned on, horizontal trajectories can recombine with the more traditional 45$^\circ$ trajectories discussed in this work \cite{Brower:2006ea}.\footnote{The terminology ``horizontal" and ``$45^\circ$" refers to the Chew-Frautschi plot, where we plot $S$ vs.\ $\De-\frac d 2$.} In perturbation theory, such recombinations manifest as poles in anomalous dimensions and structure constants, which satisfy compatibility conditions between the two combining branches, see e.g.\ \cite{Brower:2006ea,Costa:2012cb}. A class of horizontal trajectories that should recombine with the twist-3 operators considered here was explored in \cite{Korchemsky:2003rc}. It would be very interesting to describe the recombinations explicitly at the level of operators, using the techniques of \cite{Caron-Huot:2022eqs}. We also expect branch cuts in perturbation theory where different twist-3 trajectories recombine, see figure~\ref{C123complexPlane}.

Forging a path into the LHP is a difficult task. Newton's series will converge to the right of the first singularity. At that point, a wall will emerge that we cannot immediately pass through -- see figure \ref{example}. If we know the location and nature of the singularity, we can subtract it out, tear down the wall and continue further till we hit the next singularity and so on. This is the case for the integer poles -- those are easy to take care of. If we do not know precisely where these singularities are -- as for the cuts -- we need better tools. We are currently honing these weapons and hope to delve into the LHP soon and come back with messages from the dragons therein.

{\begin{center}{\textbf{ACKNOWLEDGMENTS}} \end{center}}
We thank F.~Aprile, B.~Basso, C.~Bercini, C-H.~Chang, F.~Coronado, S.~Caron-Huot, N.~Gromov, P.~Kravchuk, and V.~Voloshyna, X.~Yin for enlightening discussions. We are especially grateful to Simon Caron-Huot for a most inspiring discussion in 2018, at the annual Simons Bootstrap Collaboration  meeting at the Perimeter Institute. Some of the results obtained in this letter were somehow anticipated by Simon already at that time (albeit using different tools/ideas).
We thank several participants of the 2022 Simons bootstrap collaboration meeting for reminding us of the important reference \cite{BaltMadalena}.
Research at the Perimeter Institute is supported in part by the Government of Canada through NSERC and by the Province of Ontario through MRI. This work was additionally supported by a grant from the Simons Foundation (PV: \#488661) and FAPESP grant 2016/01343-7 and 2017/03303-1.  DSD is supported by Simons Foundation grant 488657 (Simons Collaboration on the Nonperturbative Bootstrap) and a DOE Early Career Award under grant No.~DE-SC0019085.

\appendix

\section{Structure Constants From Integrability} \label{Cappendix}

Local operators in $\mathcal{N} =4$ SYM are described by $\mathcal{Q}$-functions. For SL$(2)$ operators such as (\ref{family}) the $\mathcal{Q}$-functions reduce, at leading order, to the Baxter polynomials $Q(x) = \prod_{j=1}^S(x-u_j)$. Structure constants between such operators and BPS scalars are given by simple overlaps between the polynomials $Q$,
\beq
\frac{C(S)^2}{C_\text{BPS}^2} = \frac{(S!)^2}{(2S)!}\frac{\langle Q,\mathbf{1}\rangle_{\ell}^2}{\langle Q,Q\rangle_{L}} \label{c123sov}
\eeq
where $C_\text{BPS}^2$ is the (protected) structure constant between the 2 fixed BPS scalars and the BPS operator given by~(\ref{family}) with $S=0$. 

The case we are interested is $L=3$, $\ell=1$, corresponding to the twist 3 operators (\ref{family}). For $\ell = 1$ the numerator overlap trivializes, $\langle Q,\mathbf{1}\rangle_{\ell}^2=1$, while the \textit{norm} $\langle Q,Q\rangle_{L}$ is non-trivial. It is given in terms of the roots $\{u_1,\dots,u_S\}$ through a determinant \cite{Vieira:2013wya}
\beq
\langle Q,Q\rangle_{L} =\frac{\det\left(\partial_{u_i} \phi_j\right) Q(i/2)Q(-i/2)}{(2S)!\prod_{i\neq j}^S \frac{u_i - u_j}{u_i - u_j -i}},  \label{gaudindet}
\eeq
with $e^{i \phi_j} = (\frac{u_j + i/2}{u_j -i/2})^L\prod_{k\neq j} \frac{u_j - u_k + i}{u_j-u_k - i}$, or through the Separation of Variables integrals \cite{Derkachov:2002tf,Bercini:2022jxo,Jiang:2016ulr,Cavaglia:2019pow}
\beq
\langle Q,Q\rangle_{L} = \binom{2 S + L-1}{L-1} \int_{\mathbb{R}^{L-1}} \mu_L \prod_{i=1}^{L-1}  Q(x_i)^2  \label{sovformula},
\eeq
with factorized measure
\beq
d\mu_{L} = \prod_{i=1}^{L-1} dx_i \, \mu_1(x_i)\prod_{i=1}^{L-2}\prod_{j=i+1}^{L-1}\mu_2(x_i,x_j) \nonumber
\label{multimeasure}
\eeq
where
\beq
  \mu_1(u)\! =\! \frac{\pi}{2 \cosh(\pi u)^2} \nonumber , \,\,\,\,\,
 \mu_2(u,v)\! =\! \frac{\pi(u-v)\sinh(\pi (u-v))}{\cosh(\pi u)\cosh(\pi v)}.
 \nonumber
\eeq
Once the roots are determined through (\ref{electro}), either (\ref{gaudindet}) or (\ref{sovformula}) can be used to generate the figure \ref{twist3ope} data.

\section{Twist 3 Lowest Family} \label{lowestFamilyAppendix}
At leading order the Baxter functions are polynomials. These polynomials are known analytically in the cases of the twist 2 and lowest twist 3 trajectories. They are given~\cite{Kotikov:2008pv,Beccaria:2007cn} by
\begin{align}
\mathcal{Q}_2(u,S) &= {}_3F_2\left(1+S, -S, 1/2 + i u; 1,1|1\right), \label{twist2Q}\\
\mathcal{Q}^\texttt{lowest}_3(u,S) &= {}_4F_3\left(-\tfrac{S}{2}, 1+\tfrac{S}{2}, \tfrac{1}{2} + i u,\tfrac{1}{2} - i u; 1,1,1|1\right)\label{twist3Q}.
\end{align}

In this section we use this result to determine the structure constants for these trajectories in closed form. Note that naively these Q-functions are not appropriate to describe the trajectory at non-(even-)integer values of spin. For example, $\mathcal{Q}^\text{lowest}_3(u,S)$  has a symmetry~$S\rightarrow-2-S$ which is not a symmetry of the anomalous dimension~$\gamma_3^\text{lowest} = 8H(S/2)$. This will be reflected in the analytic structure constant obtained.

We are therefore interested in the structure constants described in (\ref{c123sov}) in the case

$L=3$ and $\ell=1$. For~$\ell=1$ the numerator in (\ref{c123sov}) trivializes since there are no integrals to be performed. Only the denominator is non-trivial. 
The SOV integrals (\ref{sovformula}) simplify in the tree-level approximation and acquire determinant form. In the case of interest we have
\beq
\langle \mathcal{Q}^\texttt{lowest}_3, \mathcal{Q}^\texttt{lowest}_3 \rangle_3 = \binom{2S+2}{2} \det \left[\int d\mu_{i,j}(u) \mathcal{Q}^\texttt{lowest}_3\right] \nonumber
\eeq
with
\beq
 d\mu_{i,j}(u) = u^{i} \pi \cosh(\pi u)^{-2} \tanh(\pi u)^{j}, \qquad i,j  = 0,1. \nn
\eeq
Moreover, since the lowest trajectory $\mathcal{Q}^\text{lowest}_3$ is an even polynomial at even spins, the off-diagonal integrals vanish. We therefore have
\begin{align}
\det \left[\int d\mu_{i,j}(u) \mathcal{Q}^\texttt{lowest}_3\right]
= \left(\mathcal{I}_1 \equiv \int du \frac{\pi}{2} \frac{ (\mathcal{Q}^\texttt{lowest}_3)^2}{\cosh(\pi u)^{2}}\right) & \nn\\ \times \left(\mathcal{I}_2 \equiv\int du \frac{\pi^2 u \tanh(\pi u)  (\mathcal{Q}^\texttt{lowest}_3)^2 }{\cosh(\pi u)^{2}}\right) & \nn.
\end{align}

The integral $ \mathcal{I}_2$ is simple. One can check that it evaluates to $ \mathcal{I}_2 =  (S +1)^{-1}$. This is shown in appendix \ref{orthoapp}. We are thus left with the evaluation of $\mathcal{I}_1$. We do so through two observations. First, note that  $\mathcal{I}_1$ measure is an orthogonal measure for the twist 2 Q-functions (see~\cite{Derkachov:2002tf,Cavaglia:2019pow,Bercini:2022jxo}):
\beq
\int du  \frac{\pi}{2} \frac{ \mathcal{Q}_2(u,S)\mathcal{Q}_2(u,S')}{\cosh(\pi u)^{2}} = \delta_{S, S'} (-1)^S (2 S+1)^{-1}\nonumber \,.
\eeq
We also derive this integral in appendix \ref{orthoapp}.
Second, note that, for even spin $S$, the Q-functions (\ref{twist2Q}, \ref{twist3Q}) are even polynomials of degree $S$. One can therefore decompose the twist 3 polynomials (\ref{twist3Q}) in a basis of twist 2 polynomials (\ref{twist2Q}). The result is
\begin{align*}
&\mathcal{Q}^\texttt{lowest}_3(u,S) = \sum_{j=0}^{S/2} \mathcal{Q}_2(u,2j) \times \\ &
\left(\frac{i^{2j + S} (1 + 4 j) \Gamma\left(\tfrac{1}{2} +  j\right)^2\Gamma\left((1+\tfrac{S}{2}+j)\right)/\Gamma\left(1+j\right)^2 }{2 \Gamma\left(\tfrac{1}{2}-\tfrac{S}{2}+j\right) \Gamma\left(1+\tfrac{S}{2}-j\right)\Gamma\left(\tfrac{3}{2} +\frac{S}{2}+j\right)} \right) 
\end{align*}
 We can therefore combine these two observations 
 to write
 \beq
 \mathcal{I}_1 = \sum_{j=0}^{S/2}\left( \frac{\sqrt{1+4j}\Gamma\left(\tfrac{1}{2} +  j\right)^2 \Gamma\left(1+\tfrac{S}{2}+j\right)/\Gamma\left(1+j\right)^2}{2  \Gamma\left(\tfrac{1}{2}-\tfrac{S}{2}+j\right) \Gamma\left(1+\tfrac{S}{2}-j\right)\Gamma\left(\tfrac{3}{2} +\tfrac{S}{2}+j\right)} \right)^{2}, \label{sumfinal}
 \eeq
The sum (\ref{sumfinal}) can be extended to infinity.  Combining with (\ref{sovformula}) leads to the final result (\ref{C123lowest}) which holds for any even integer $S$. 

\subsection{Hypergeometric integrals, orthogonality and recursions}
\label{orthoapp}
In this subsection we compute the integrals
\begin{align}
&\mathcal{I}_A(S)= \int du \frac{\pi}{2} \frac{ \mathcal{Q}_2(S)^2}{\cosh(\pi u)^{2}}, \label{tocompute1}\\ & \mathcal{I}_{B}(S) = \int du \frac{\pi^2 u \tanh(\pi u)  \mathcal{Q}^\text{lowest}_3(S)^2 }{\cosh(\pi u)^{2}}. \label{tocompute2}
\end{align}
Key are the recursion relations satisfied by the Hahn polynomials (\ref{twist2Q},\ref{twist3Q}),
\begin{align}
&(S+2)^2\mathcal{Q}_2^{[2]} + 2 i (3 + 2 S) u \mathcal{Q}_2^{[1]} = (S+1)^2 \mathcal{Q}^{[0]}_2,  \label{rec2}\\
&(S+2)^3\mathcal{Q}^{[2]}_3 + S^3\mathcal{Q}^{[-2]}_3 = 2(1+S)(2 + 2S+S^2-8 u^2)\mathcal{Q}^{[0]}_3. \label{rec3}
\end{align}
with $\mathcal{Q}^{[a]}_2 \equiv \mathcal{Q}_2(S+a)$ and $\mathcal{Q}^{[a]}_3 \equiv \mathcal{Q}^{\texttt{lowest}}_3(S+a)$. As follows from the SOV methods of \cite{Bercini:2022jxo}, the kernels of~(\ref{tocompute1},~\ref{tocompute2}) define orthogonal scalar products for the trajectories under consideration:
\begin{align}
&\int du \frac{\pi}{2} \frac{ \mathcal{Q}_2(S) \mathcal{Q}_2(S')}{\cosh(\pi u)^{2}} \propto \delta_{S S'}, \label{ortho2} \\& \int du \frac{\pi^2 u \tanh(\pi u)  
\mathcal{Q}^\texttt{lowest}_3(S)\mathcal{Q}^\texttt{lowest}_3(S') }{\cosh(\pi u)^{2}} \propto \delta_{S S'}. \label{ortho3}
\end{align}

To compute $\mathcal{I}_A(S)$, integrate equation (\ref{rec2}) against $\mathcal{Q}_2(S) \tfrac{\pi}{2} \cosh(\pi u)^{-2}$. 
Using the decomposition
\beq
u \mathcal{Q}_2(S) =  i \frac{(1+S)^2}{2+4 S}\mathcal{Q}_2(S+1) + \sum_{k=0}^{S}  c_k \mathcal{Q}_2(k), \nonumber
\eeq
which follows from matching asymptotics, and the orthogonality relation (\ref{ortho2}), we obtain the recursion
\beq
\frac{3+2S}{1+2S} \,\mathcal{I}_A(S+1) +\mathcal{I}_A(S) =0 \nonumber
\eeq
whose solution is 
\beq
\mathcal{I}_A(S) = \frac{(-1)^S}{2S+1}\,. \la{guess1}
\eeq

Similarly, to compute $\mathcal{I}_B(S)$, integrate (\ref{rec3}) against $\pi^2 u \tanh(\pi u)  \mathcal{Q}^\texttt{lowest}_3(S-2) \cosh(\pi u)^{-2}$. Using the decomposition
\beq
u^2 \mathcal{Q}^\texttt{lowest}_3(S-2) =  \tfrac{16 (1-S)}{S^3} \mathcal{Q}^\texttt{lowest}_3(S) + \sum_{k=0}^{\frac{S}{2}-1}  c_{2k} \mathcal{Q}^\texttt{lowest}_3(2k) \nonumber
\eeq
and the orthogonality relation (\ref{ortho3}) we obtain
\beq
\frac{1+S}{1-S}\, \mathcal{I}_B(S) + \mathcal{I}_B(S-2)=0 \nonumber
\eeq
from which follows
\beq
\mathcal{I}_B(S) = \frac{1}{S+1}\,. \la{guess2}
\eeq
The derivation in this section is a bit of an overkill given the simplicity of the final results. It would be much simpler to evaluate the integrals (\ref{tocompute1}) and (\ref{tocompute2}) for the first few physical spins $S$ and immediately recognize (\ref{guess1}) and (\ref{guess2}). (This was of course how we first found them.)

\section{Baxter at Complex Spin}
\la{baxterAppendix}
Analysis of the asymptotics of (\ref{baxterEq}) determine that the leading power-law behaviour of $Q$ can be either~$u^S$ or~$u^{-2-S}$. In \cite{Janik:2013nqa} Janik proposed, in the case of twist 2 operators, 
that it is the second class of solutions which control the correct analytic continuation in spin of the physical data. Note: as one approach integer spin,~$Q$ must approach point-wise the polynomial solutions with~$u^S$ asymptotics describing the local operators while decaying (in the RHP) as $u^{-2-S}$ when $u \rightarrow + \infty$! Once the decaying $Q$ is known, the energy can be extracted from
\beq
\gamma = 2 i \oint \frac{du}{2\pi i} \frac{1}{u^2}\log\left(\frac{Q(u+i/2)}{Q(u-i/2)}\right) \la{newenergy},
\eeq
which generalize (\ref{energy}) to the non-polynomial case.

Baxter equation, being a finite difference equation, admits a gauge redundancy $Q(u) \rightarrow Q(u) p(u)$ for any i-periodic $p$. The energy is invariant under this transformation\footnote{The asymptotics condition can also be expressed in an invariant form as the requirement that $\log\left(Q(u+i)/Q(u-i)\right) \sim i (M+2)/u$ as $u \rightarrow \infty$.}.  This gauge invariance is crucial for the method proposed here. Solutions with leading power-law asymptotics have poles at $u= \frac{i}{2} + i k$, $k \in \mathbb{Z}$. These poles can be removed through multiplication by  $\sinh(2\pi u)$ factors so that we are left with an entire solution normalizable at~$i/2$. This does not fix the gauge freedom completely as we can still multiply $Q$ by, say, $\cosh(\pi u)^2$ factors. These would increase the exponential rate of $Q$ as $u \rightarrow \infty$. We therefore look for the slowest growing entire solutions to~(\ref{baxterEq}). In section \ref{BaxterBlessing} we propose a numerical algorithm that determines this solution and computes the correct continuation of the energies.

\section{Twist 2, Newton Series and Integrability} \label{twist2}

The analytically solvable twist 2 trajectory 
\beq
\Tr\left(Z D^S_+Z\right) + \texttt{permutations}
\eeq
serves as the perfect toy model to test the ideas presented in this letter. 
Its anomalous dimension~$\gamma_2(S) = 8 H(S)$ and structure constant (with two BPS operators)~$C(S)^2_{\texttt{twist-2} LO} = \frac{2(S!)^2}{(2S)!}$ can be extracted from the four point correlator and are given by \cite{Dolan:2000ut,Dolan:2003hv}.

Alternatively, it can be computed from integrability through the twist 2 Baxter equation
\begin{align}
(u+i/2)^2 Q(u+i) +& (u-i/2)^2 Q(u-i)\nonumber\\ = &(2 u^2 - S (S+1)-1/2)Q(u), \nonumber
\end{align}
whose solution at integer spin is given by Hahn polynomials $\mathcal{Q}_2(u,S)$, equation (\ref{twist2Q}).

Newton series can be used to reproduce $\gamma_2$ and~$C(S)^2_{\texttt{twist-2} LO}$ everywhere in the complex plane provided one subtracts singularities as described in figure~\ref{example}. In the case of the structure constant it is important to subtract the exponential behaviour through multiplication by $e^{\left(2 \log(2) S\right)}$ before applying the interpolation method.

One can also extract the energies directly at complex spin, as proposed by Janik in \cite{Janik:2013nqa}, through the slow-growing solution described in section \ref{BaxterBlessing}. In this case the slow-growing Q-function can be written analytically as 
\begin{align}
\mathcal{Q}_\texttt{slow}=&\frac{i\sinh(2 \pi u)}{2\sin(\pi S)} \Bigg(\left(1 - i \tan(\pi \tfrac{S}{2}) \coth(\pi \nonumber u)\right)\mathcal{Q}_2(u,S)-\\&\left(1 + i \tan(\pi \tfrac{S}{2}) \coth(\pi u)\right)\mathcal{Q}_2(-u,S)\Bigg)
\end{align}
from which one reads the energy
\beq
\gamma = 2 i \left(\frac{\mathcal{Q}_\texttt{slow}'(\tfrac{i}{2})}{\mathcal{Q}_\texttt{slow}(\tfrac{i}{2})} - \frac{\mathcal{Q}_\texttt{slow}'(-\tfrac{i}{2})}{\mathcal{Q}_\texttt{slow}(-\tfrac{i}{2})} \right) = 8 H(S).
\eeq

As mentioned in section \ref{discussion} there are currently no available integrability methods to extract structure constants directly at complex spin. On the other hand, in the case of twist 2 the light-ray operators that physically realize the complex spin data $\{\gamma_2,C(S)^2_{\texttt{twist-2} LO}\}$ can be constructed explicitly \cite{Kravchuk:2018htv}, providing a direct method to extract the data. This is reviewed in section \ref{lightray}.

\section{Twist 2 and Light-ray operators}
\label{lightray} In this appendix we review how the matrix elements of the explicit leading order twist 2 light-ray operators constructed in \cite{Kravchuk:2018htv} encode the complex spin structure constants. 

The leading order twist 2 local (even $S$) primary operators are given by
\beq
\mathcal{O}_{S}(x) = \frac{1}{\sqrt{2 (2S)!}}\sum_i \psi_i \Tr\left(D_+^i Z D_+^{S-i} Z \right)(x),\nonumber
\eeq
where $\psi_i =  (-1)^i \binom{S}{i}^2$ is completely fixed by the primary condition $[K_-,O_{S}(0)] = 0$. The operators are unit-normalized so that
\begin{equation}
\langle \bar{\mathcal{O}}_{S}(x)  {\mathcal{O}}_{S}(0) \rangle = \frac{( x^-)^{2S}}{(x^2)^\Delta}. \nonumber
\end{equation}
The three point function between $\mathcal{O}_{S}$ and $\tfrac{1}{2}$-BPS operators \footnote{Explicitly, consider $O^{(2)}= \Tr(\bar{Z}\bar{X})$ and $ O^{(3)}= \Tr(\bar{Z}X) $.
} $O^{(2)}$ and $O^{(3)}$ is then given by
\beq
\langle O^{(2)}(x_2) \mathcal{O}_S(x_1) O^{(3)}(x_3) \rangle = \frac{\sqrt{2}(S!)}{\sqrt{(2S)!}} \frac{\left(  x_{13}^- x_{12}^2 -  x_{12}^- x_{13}^2  \right)^S }{x_{12}^{2+2S} x_{13}^{2+2S}x_{23}^2} \label{3ptlocal},
\eeq
\linebreak
from which one reads $C(S)^2 = \tfrac{2(S!)^2}{(2S)!}$. We assumed the insertions were space-like separated. 

The operators $\mathcal{O}_S$ only make sense at integer spin. Our goal is to construct operators in continuous spin representations whose matrix-elements at integer $S$ reproduce the local operator data $C(S)$. These are necessarily non-local \cite{Kravchuk:2018htv,Mack:1975je}.  With that purpose in mind, consider the light-transform of operator $\mathcal{O}_S$
\beq
L[\mathcal{O}_S] = \int\limits_{-\infty}^{\infty}d\alpha  \,\mathcal{O}_S(\alpha n^+). \label{lighttransform}
\eeq
In \cite{Kravchuk:2018htv} it was shown that these operators transform as primaries with dimension $\Delta_L = 1-S$ and spin $S_L = 1-\Delta$ inserted at past-null infinity (at $(0,-\infty,0)$ in light-cone coordinates $(x_-,x_+,\vec{x})$). 

Indeed, fixing the insertions as in figure \ref{lightrayfigure} and acting with the transform on (\ref{3ptlocal}) while being careful with implicit $i\epsilon$s we obtain\footnote{As usual, when inserting the operator at past null infinity we rescale the correlator by a factor $(x_1^+)^{\Delta_L + S_L}$ before taking the $x_1^+ \rightarrow\infty$ limit.}
\begin{equation}
\langle \Omega| O^{(2)} L[\mathcal{O}_S] O^{(3)} | \Omega \rangle  =\frac{C(S) f(S) }{x_{2}^- x_{3}^- x_{23}^2} \left(  \frac{x_{3}^2}{x_3^-} - \frac{x_{2}^2}{x_{2}^-}  \right)^{-1-S}  \label{structurecomplex},
\end{equation}
with $f(S) = 2 \pi i  \Gamma(2S+1)/\Gamma(S+1)^2$. This is precisely the structure of a correlator of two BPS scalars and a primary with quantum numbers $(\Delta_L,S_L)$. 

So far, all we did was to perform an integral transform. However, we are rewarded once we realize (\ref{lighttransform}) belong to a continuous family of \textit{light-ray} operators
\begin{align}
& \mathbb{O}(S) = \frac{i}{4\pi} \frac{\sqrt{2} \sqrt{\Gamma(2 S+ 1)}}{\Gamma(S+1)} \times \label{lightrayexplicit} \\& \int_{-\infty}^{\infty}d\alpha d\beta \left(\frac{1}{(\alpha - \beta + i \epsilon)^{S+1}}+ (\alpha \leftrightarrow \beta)\right) \Tr\left(Z(\alpha) Z(\beta)\right) \nonumber.
\end{align}
which transform as primaries with quantum numbers $(\Delta_L,S_L)$ for \textit{arbitrary} values of $S$ \cite{Kravchuk:2018htv}! Indeed, at even $S$ we have
\beq
\left(\frac{1}{(\alpha - \beta + i \epsilon)^{S+1}}+ (\alpha \leftrightarrow \beta)\right) = - \frac{2 \pi i}{\Gamma(S+1)} \delta^{(S)}(\alpha-\beta) \nonumber,
\eeq
so that in this case
\beq
\mathbb{O}(S) = \frac{\sqrt{\Gamma(2S+1)}}{\sqrt{2} \Gamma(S+1)^2} \int d\alpha  \Tr\left(Z\mathcal{D}^S_+Z\right)(\alpha) = L[\mathcal{O}_S] \nonumber.
\eeq
where we used integration by parts to act with all derivatives inside $\mathcal{O}_S$ on the second field. Hence, at the even integers the partons must move together along the light-ray and we recover the (correctly normalized) light-transform of the local operator. 

\begin{figure}[t]
  \includegraphics[width=0.5\textwidth]{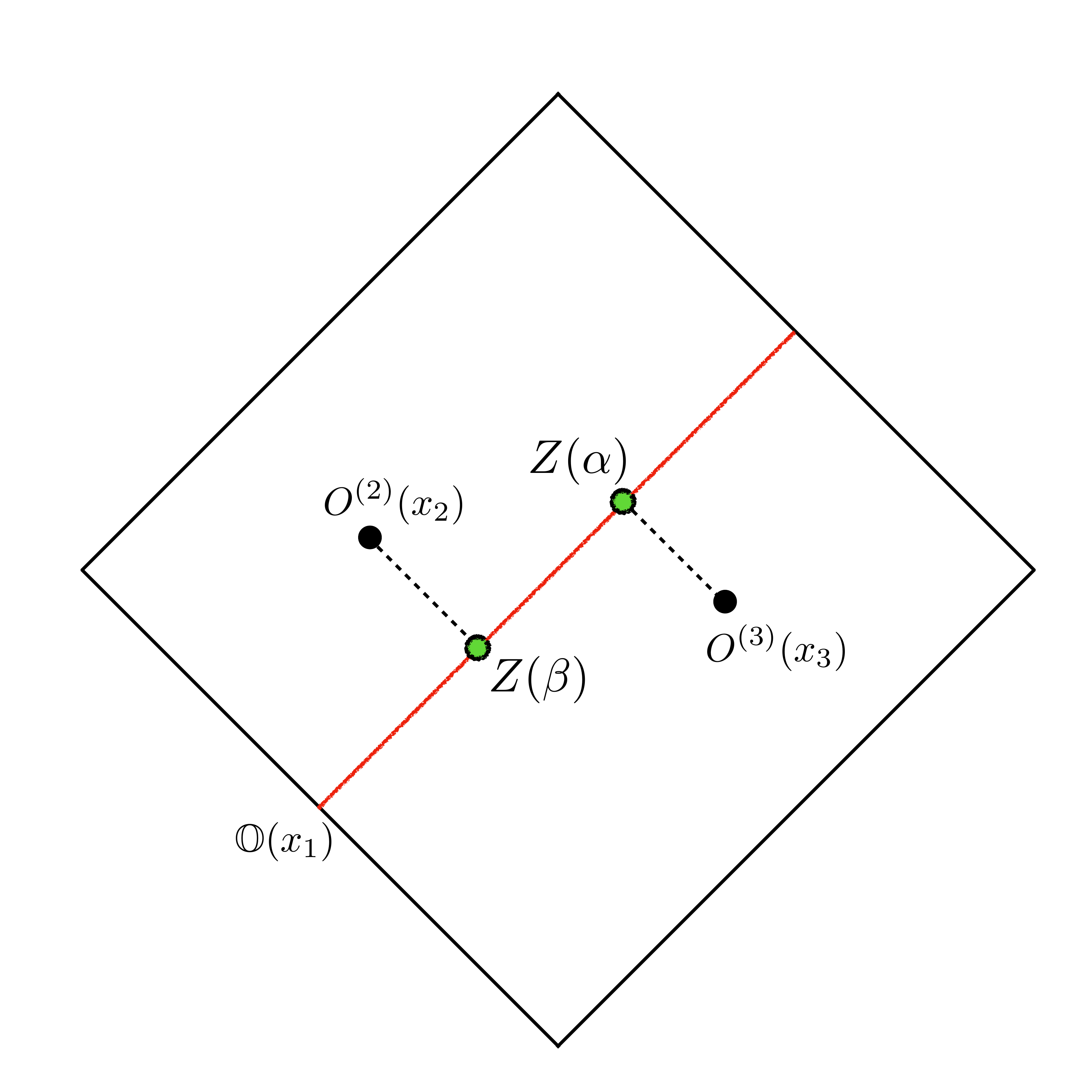}
  \caption{We insert the operators at $x_1^-=0$, $x_2^->0$ and $x_3^-<0$. When computing the matrix element $\langle \Omega| O^{(2)} \mathbb{O}_S O^{(3)} | \Omega \rangle $ the green partons are integrated along the light-ray, equation (\ref{lightrayintegral}). The integral is evaluated by picking residues that localize the partons on the null-cone of the BPS insertions.} \la{lightrayfigure}
\end{figure}

For non-integer $S$, $\mathbb{O}$ provides a physical realization of the structures $C(S)$. To see that, compute through Wick contractions the matrix element
\begin{align}
&\langle \Omega| O^{(2)} \mathbb{O}_S O^{(3)} | \Omega \rangle = \frac{i}{4\pi} \frac{\sqrt{2}\sqrt{\Gamma(2 S+ 1)}}{\Gamma(S+1)} \int \frac{d\alpha d\beta }{(\alpha - \beta + i \epsilon)^{S+1}}   \nonumber \\ & \times \frac{2}{\alpha x_2^- + x_2^2 - i \epsilon}\frac{1}{\beta x_3^- + x_3^2 - i \epsilon} \frac{1}{x_{23}^2} + (\beta \leftrightarrow \alpha) \label{lightrayintegral}.
\end{align}
The $i \epsilon$s are crucial and follow from the operator ordering. Since $x_2^- >0$ and $x_3^-<0$, see figure \ref{lightrayfigure}, only the first term contributes since otherwise the contours can be deformed to infinity and the integral vanishes. For the first term, the integral in pinched by the singularities. Picking the residues from the propagators, see figure \ref{lightrayfigure}, we obtain
\begin{equation}
\langle \Omega| O^{(2)} \mathbb{O}_S O^{(3)} | \Omega \rangle =  \frac{f(S) C(S)}{x_2^- x_3^- x_{23}^2} \left(\frac{x_3^2}{x_3^-}-\frac{x_2^2}{x_2^-}\right)^{-S-1}\nonumber,
\end{equation}
which match (\ref{structurecomplex}) exactly.

The anomalous dimension can be read off similarly.
The one-loop dilatation action on the Wilson line insertions is given by \cite{Sever:2012qp,Belitsky:2004cz,Belitsky:2011nn}
\begin{align}
&\mathcal{D}\circ \Tr\left(Z(\alpha) Z(\beta)\right) =4g^2 \int_0^1 \frac{d\tau}{\tau} \nonumber \Bigg( \Tr\left(Z(\alpha) Z(\beta)\right) \\&-\Tr\left(Z(\alpha (1-\tau) + \beta \tau) Z(\beta)\right) + (\alpha \leftrightarrow\beta)\Bigg).
\end{align}
Acting with $\mathcal{D}$ on (\ref{lightrayexplicit}) and changing variables to transpose the convolution from the fields to the wavefunction, we readily obtain
\beq
\mathcal{D} \circ \mathbb{O}(S) = \left(8g^2\underbrace{\int_0^1 \frac{d\tau}{\tau} \left(1-(1-\tau)^S\right)}_{H(S)} \right) \mathbb{O}(S) \nonumber.
\eeq

Thus, we are able to reconstruct the complex spin data $\{\gamma(S), C(S)\}$ from the explicit wavefunctions of the light-ray operators. In the twist 2 case the computation is sort of trivial: $C(S)$, for example, is essentially encoded in the normalization of the operator so that when computing the matrix elements all the integrals do is reproduce the correct tensor structures. This is not a surprise as both the twist 2 local operators and light-rays have their wavefunctions completely fixed by conformal symmetry.

It is an important open problem to generalize this construction to higher-twist operators. There we do not expect such trivialities to occur, and the structure of the wavefunction should contribute non-trivially to the complex spin structure constants. Besides giving direct access to the complex spin data, such construction should clarify a number of puzzles regarding the higher-twist trajectories. See further discussion in section \ref{discussion}.

\section{Hydrogen Atom} \label{hydrogenAp}

When we presented this work at the 2023 Annual Bootstrap meeting at the Simons Foundation, Simon Caron-Huot suggested that decoupling zeroes might be present in the literature in relation to  the hydrogen atom, the model which gave rise to the development of Regge theory \cite{Regge:1959mz}. We did not find  such reference(s) but it is indeed very instructive to see that similar decoupling zeroes are already present in the exactly solvable Coulomb potential scattering problem, as described in this appendix. 

\begin{figure}[t]
  \includegraphics[width=0.5\textwidth]{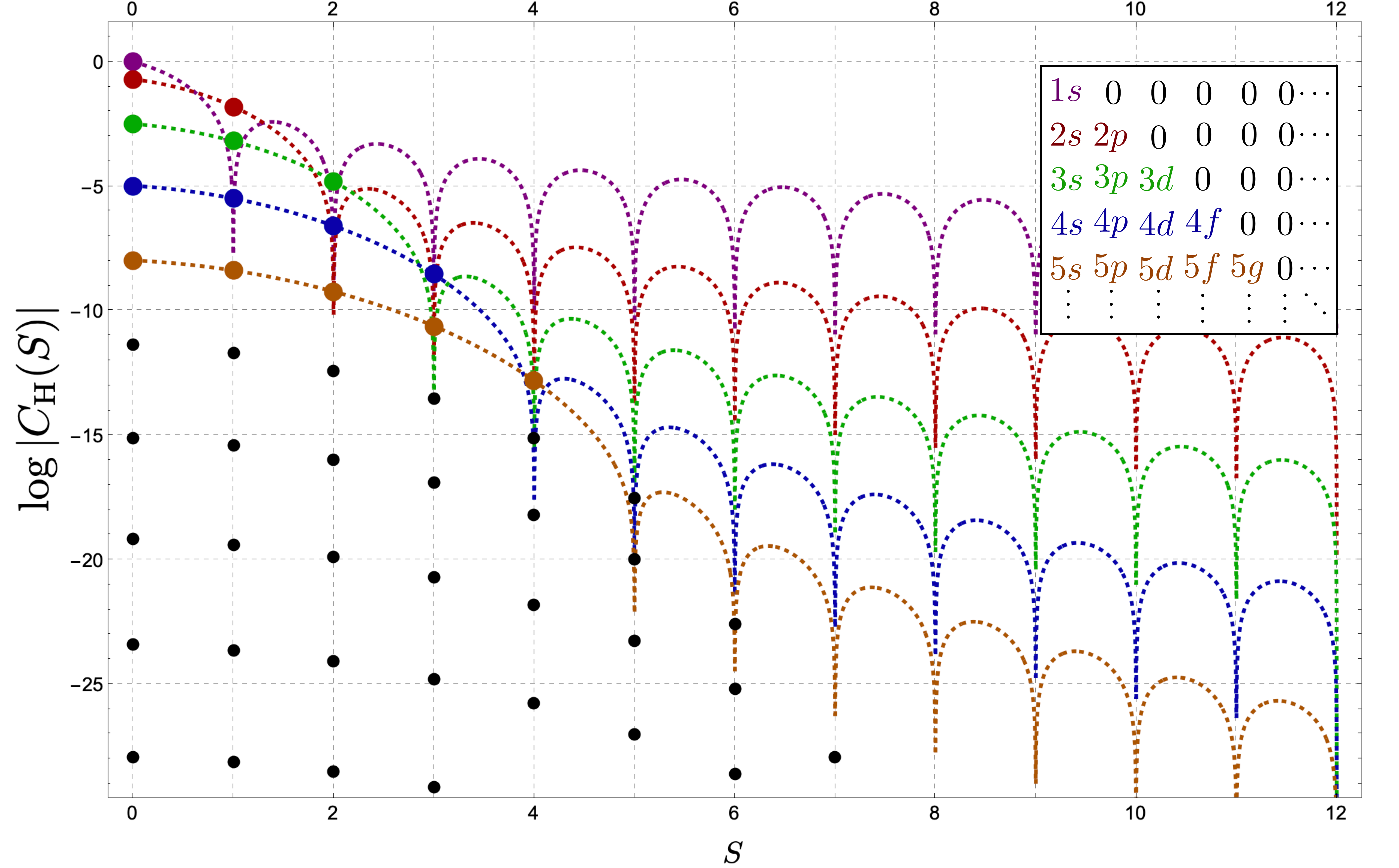}
  \caption{Hydrogen atom coupling constants (\ref{structureHydrogen}) as a function of spin $S$ for the various energy levels $n$. Dots correspond to the physical states, see also inset. We colour the first five trajectories. For large spin there are no corresponding phyiscal states at integer $S$ and the coupling constant develops decoupling zeroes.} \la{hydrogenfigure}
\end{figure} 

Consider the scattering of an electron against a proton in the Coulomb approximation. The spin $S$ partial-waves 
\beq
f_S(E) = \frac{\Gamma(1+S- \eta)}{\Gamma(1+S+ \eta)}, \nonumber
\eeq
where $\eta$ is related to the energy as $\eta = 1/\sqrt{-E}$ in our conventions. The partial-wave $f_S$ has poles when the scattering energy (or $\eta$) equals the (negative) binding energy (or principal quantum number $n$) of a bound-state with spin $S$. Recall that bound states must have~$S + 1 \leq n$.
Define the ``coupling constant" between the scattering state and the bound-state with principal quantum number $n$ and spin $S$ as
\beq
\left|C^\text{H}_n(S)\right| = \left|\underset{\eta = n}{\text{res}} f_S(\eta)\right| = \frac{1}{\Gamma(n-S)\Gamma(1+n+S)} \label{structureHydrogen}.
\eeq

As in the main text, we group states through the ordering of their energies at fixed $S$, i.e families are labelled by $n$. The growth of states here is, however, the opposite of the CFT: in the hydrogen atom states disappear as we keep S fixed and decrease the energy, while in the twist-3 sector of the CFT states disappear at high dimension. Even though families have finitely many states, the coupling constants nevertheless organize themselves into analytic functions of $S$, (\ref{structureHydrogen}). Moreover, at large values of $S$ for which there are no physical states the coupling constant (\ref{structureHydrogen}) vanishes, see figure (\ref{hydrogenfigure}). This is equivalent to the decoupling mechanism of figure (\ref{summaryallfamilies}).

Note, however, that in the case of the Hydrogen atom there are many other ways of grouping states in analytic families. For example, one could consider diagonal trajectories $n-S = \text{C} \in \mathbb{N}$. Each of these families would contain operators of all spins. Correspondingly, no decoupling mechanism would be present and the coupling constants (\ref{structureHydrogen}) at fixed $C$ would have no zeroes in the RHP of $S$.

\bibliography{refs}

\end{document}